\documentclass[fleqn,useAMS,usenatbib]{mnras}

\usepackage{newtxtext,newtxmath}
\usepackage[T1]{fontenc}
\usepackage{ae,aecompl}

\usepackage{graphicx}	% Including figure files
\usepackage{amsmath}	% Advanced maths commands
\usepackage{amssymb}	% Extra maths symbols
\usepackage{booktabs}
\usepackage{url}
\usepackage{multirow}
\usepackage{fix-cm}     %remove the warning of font size

\newcommand{\hmpc}{h^{-1}\,{\rm Mpc}}
\newcommand{\patchy}{{\textsc{Patchy}}}
\newcommand{\lowz}{low-$z$}
\newcommand{\highz}{high-$z$}

\title[BAO with voids and galaxies]{Improving baryon acoustic oscillation measurement with the combination of cosmic voids and galaxies}

\author[C. Zhao et al.]{
\parbox{\textwidth}{
Cheng Zhao,$^{1,2,3}$\thanks{E-mail: cheng.zhao@epfl.ch}
Chia-Hsun Chuang,$^{4,5}$\thanks{E-mail: chuangch@stanford.edu}
Francisco-Shu Kitaura,$^{6,7}$
Yu Liang,$^2$
Marcos Pellejero-Ibanez,$^{6,7}$
Charling Tao,$^{8,2}$
Mariana Vargas-Maga\~na,$^{9}$
Andrei Variu,$^1$
Gustavo Yepes$^{10}$
}
\\
\vspace*{4pt} \\
$^{1}$Laboratory of Astrophysics, Ecole Polytechnique F\'ed\'erale de Lausanne (EPFL), Observatoire de Sauverny, 1290 Versoix, Switzerland\\
$^{2}$Tsinghua Center for Astrophysics (THCA) \& Department of Physics, Tsinghua University, Beijing 100084, P.~R.~China\\
$^{3}$National Astronomical Observatories, Chinese Academy of Science, Beijing, 100012, P.~R.~China\\
$^{4}$Kavli Institute for Particle Astrophysics and Cosmology \& Physics Department, Stanford University, Stanford, CA 94305, USA\\
$^{5}$Leibniz-Institut f\"{u}r Astrophysik Potsdam (AIP), An der Sternwarte 16, D-14482 Potsdam, Germany \\
$^{6}$Instituto de Astrof\'{\i}sica de Canarias (IAC), C/V\'{\i}a L\'actea, s/n, E-38200, La Laguna, Tenerife, Spain \\
$^{7}$Departamento Astrof\'{\i}sica, Universidad de La Laguna (ULL), E-38206 La Laguna, Tenerife, Spain \\
$^{8}$CPPM, Universit\'{e} Aix-Marseille, CNRS/IN2P3, Case 907, 13288 Marseille Cedex 9, France \\
$^{9}$Instituto de F\'isica, Universidad Nacional Aut\'onoma de M\'exico, Apdo. Postal 20-364, M\'exico \\
$^{10}$Departamento de F\'isica Te\'{o}rica and CIAFF, M\'{o}dulo 8, Facultad de Ciencias, Universidad Aut\'{o}noma de Madrid, 28049 Madrid, Spain
}

\date{Accepted XXX. Received YYY; in original form ZZZ}

\pubyear{2019}

\begin{document}
\label{firstpage}
\pagerange{\pageref{firstpage}--\pageref{lastpage}}
\maketitle

% Abstract of the paper
\begin{abstract}
We develop a methodology to optimise the measurement of Baryon Acoustic Oscillation (BAO) from a given galaxy sample. In our previous work, we demonstrated that one can measure BAO from tracers in under-dense regions (voids). In this study, we combine the over-dense and under-dense tracers (galaxies \& voids) to obtain better constraints on the BAO scale.
To this end, we modify the de-wiggled BAO model with an additional parameter to describe both the BAO peak and the underlying exclusion pattern of void 2PCFs. We show that after applying BAO reconstruction to galaxies, the BAO peak scale of both galaxies and voids are unbiased using the modified model. Furthermore, we use a new 2PCF estimator for a multi-tracer analysis with galaxies and voids.
In simulations, the joint sample improves by about 10\,\% the constraint for the post-reconstruction BAO peak position compared to the result from galaxies alone, which is equivalent to an enlargement of the survey volume by 20\,\%.
Applying this method to the BOSS DR12 data, we have an 18\,\% improvement for the \lowz{} sample ($0.2 < z < 0.5$), but a worse constraint for the \highz{} sample ($0.5 < z < 0.75$), which is consistent with statistical fluctuations for the current survey volume.
Future larger samples will give more robust improvements due to less statistical fluctuations.
\end{abstract}

% Select between one and six entries from the list of approved keywords.
% Don't make up new ones.

%%%%%%%%%%%%%%%%%%%%%%%%%%%%%%%%%%%%%%%%%%%%%%%%%%

%%%%%%%%%%%%%%%%% BODY OF PAPER %%%%%%%%%%%%%%%%%%

\section{Introduction}

The Baryon Acoustic Oscillation (BAO) signature imprinted in the Cosmic Microwave Background (CMB) and in the three dimensional matter distribution at later cosmic times defines a characteristic scale, which is commonly used as a standard ruler to determine the expansion of the Universe \citep[][]{Blake2003, Seo2005}.
It can be measured as a peak in the 2-point correlation function (2PCF) of  matter tracers, such as galaxies \citep[][]{Eisenstein2005, Cole2005}, quasars \citep[][]{Ata2018}, the Lyman-$\alpha$ forests \citep[][]{Busca2013}, and potentially the 21\,cm line with future experiments \cite[e.g.][]{Chang2008}. 

To extract the maximum cosmological information from BAO measurements, many systematical uncertainties have to be taken into account, such as survey geometry, galaxy bias, redshift space distortions and gravitational evolution. These aspects have been addressed in the last few years to improve the BAO measurements from galaxy surveys \citep[cf.][and references therein]{Ross2017}.
In particular, ways to linearise the galaxy distribution with so-called BAO reconstruction techniques that enhance the BAO peak, have been developed and successfully applied \citep[][]{Weinberg1992, Eisenstein2007, Noh2009, Padmanabhan2012, Burden2014}.

It is important to stress that no matter how much a galaxy distribution is linearised by redistributing the position of individual galaxies according to the inferred displacement field, they will trace only the peaks of the density field, leaving a large fraction of the cosmic web under-represented, according to the galaxy bias picture introduced by \citet[][]{Kaiser1984}, which is particularly confirmed for Luminous Red Galaxies (LRGs) \citep[cf. e.g.][]{Kitaura2015b, Kitaura2016b}.
%Thus, the linearised galaxy distribution after BAO reconstruction represents at best a truncated Gaussian density field.

The three dimensional distribution of galaxies encodes also the information of local minima (troughs\footnote{`trough' refers to 3-D local minimum of the density field in this paper.}) in the density field, i.e. the distribution of under-densities where highly luminous galaxies cannot be observed, and only faint galaxies are present, due to the weakness of gravity, yielding empty regions called cosmic voids.
In particular, it is possible to estimate the depth of such cosmic voids based on the distribution of peaks (galaxies) in the density field, and thus evaluating the density field in regions with no observed galaxies.
One could use some Bayesian reconstruction techniques to achieve this \citep[cf. e.g.][]{Zaroubi1995}.
The problem of such approaches is the dependence on the cosmology set in the assumed correlation function, which we want to avoid here (although there are some attempts to do this by sampling the power spectrum, cf. \citet[][]{Kitaura2008}).
An alternative is to use pure geometrical arguments to find large regions devoid of galaxies \citep[e.g.][]{ElAd1997, Way2015, Zhao2016}, with voids being additional tracers.
The combination of this new set of cosmic void tracers with the galaxy distribution permits us to perform a multi-tracer analysis, which has been shown to yield tighter cosmological constraints \citep[e.g.][]{GilMarin2010, Bernstein2011, Hamaus2011, Hamaus2012, Abramo2013, Ferraro2015}.

Even though the exact void definition is subtle \citep[cf. e.g.][and references therein]{Colberg2008, Zhao2016}, it is widely accepted that large voids, or the so called voids-in-voids \citep[][]{Sheth2004}, are tracing troughs of the cosmic density field \citep[for a review, cf.][]{Weygaert2011}, and have negative bias \citep[e.g.][]{Hamaus2014, Zhao2016}.
Recently, it has been shown that with a definition based on Delaunay Triangulation (DT), a special type of cosmic voids (DT voids) can be used to measure the BAO signature \citep[cf.][]{Kitaura2016, Liang2016}.
DT voids that are allowed to overlap with each other, yield not only a larger sample size, but also a more detailed description of substructures of the density field. In fact, the distribution of DT void centres defines the topology of the cosmic web, and in particular, the size of large DT voids statistically represent the deepness of the under-dense regions \citep[][]{Zhao2016}. This turns out to be crucial, as the 2PCF of disjoint voids does not present a clear BAO peak \citep[][]{Kitaura2016}.
Indeed, excluding overlapping voids highly reduces the number of tracers and yields a much lower statistical probability for the detection of the BAO peak. Besides, the cosmic web is not fully described by disjoint voids, and information from the density field is lost.

%Furthermore, recently \citet[][]{Liang2016} and \citet[][]{Kitaura2016} have reported the BAO peak detection from the clustering of voids in both simulations and observational data, using the voids defined by \citet[][]{Zhao2016}. The question that whether voids can contribute to the multi-tracer constraint of cosmology naturally arises. In this paper, we aim at answering this question by examining the BAO peak position constraint of the BOSS DR12 galaxy sample with and without voids.
%On the other hand, \citet[][]{Chuang2017} has shown that anisotropic clustering model of voids is non-trivial. We cannot simply exploit the model for galaxies, since voids are indirect tracers. Therefore, we focus in this paper only isotropic BAO peak position constraints.

Since voids are indirect tracers constructed from the galaxy sample, they are strongly correlated with galaxies. Thus, an important question is how much improvement voids contribute in a multi-tracer analysis and the aim of this paper.
We describe a method to combine DT voids and galaxy distributions, and estimate the improvement on cosmological parameter estimation after applying BAO reconstruction.
In section~\ref{sec:data}, we begin with a description of the observed and simulated data. Then, we present our method of constructing void catalogues and computing 2PCFs as well as the covariances in section~\ref{sec:method}. Later, in section~\ref{sec:fit}, we develop a modified BAO fitting method. Finally, we measure the BAO peak scale by combining galaxies and voids and discuss the contribution of voids in section~\ref{sec:res}, and conclude in section~\ref{sec:con}.

%%%%%%%%%%%%%%%%%%%%%%

\section{Data}
\label{sec:data}

In this work we rely on both observed and simulated data as described below.

\subsection{Observed Galaxy sample}

We focus on the distribution of Luminous Red Galaxies (LRGs), obtained by the Sloan Digital Sky Survey (SDSS) 2.5-meter telescope at Apache Point Observatory, and presented in the final BOSS data release (DR12).
 %sample from the Data Release 12 (DR12) of the BOSS survey \citep[][]{Eisenstein2011, Dawson2013}, which uses the Sloan Digital Sky Survey (SDSS) 2.5-meter telescope at Apache Point Observatory.
 %The spectra are measured using the double-armed BOSS spectrograph, and then classified to measure the redshift \citep[][]{Bolton2012}, followed by a target selection scheme to generate the final large-scale structure catalogues \citep[][]{Reid2016}.
The LRG catalogue for large-scale structure analysis consists of over 1.3 million LRGs from both north and south galactic caps, spanning across nearly $10,000\,{\rm deg}^2$ on the sky \citep[][]{Alam2017}. We apply weights to correct for systematical effects, such as fibre collisions and the correlations between target density and both stellar density and seeing \citep[][]{Reid2016, Ross2017}, together with the redshift-dependent FKP weight for sample combination \citep[][]{FKP1994}.

As in \citet[][]{Alam2017}, we further divide the BOSS DR12 LRG sample into two independent redshift bins with nearly equal effective volume, $0.2 < z < 0.5$ and $0.5 < z < 0.75$, denoted hereafter `\lowz' and `\highz' bins respectively. The corresponding `effective' redshift for these two bins are 0.38 and 0.61.

\subsection{Simulated PATCHY halo mocks}
\label{sec:patchy_halo_mock}
To validate our BAO fitting method, we rely on 100 realisations of cubic mock halo catalogues constructed by the PerturbAtion Theory Catalogue generator of Halo and galaxY distributions \citep[the \patchy{}-code,][]{Kitaura2014}, which uses the Augmented Lagrangian Perturbation Theory \citep[ALPT,][]{Kitaura2013} to generate the dark matter density field, and then populate haloes with an explicit Eulerian non-linear and stochastic bias description.

The side length of the mock box is $2.5\,h^{-1}{\rm Gpc}$, with 960 grids on each side (i.e., $960^3$ dark matter particles in total). The \patchy{} mock halo catalogues are then generated using the public input parameters \citep[][]{Kitaura2015b} that have been calibrated with the Spherical Overdensity (SO) halo catalogue, with a halo number density of $3.5 \times 10^{-4}\,h^3\,{\rm Mpc}^{-3}$, from the BigMultiDark (\textsc{BigMD}) Planck $N$-body simulations \citep[][]{Klypin2016} at redshift $z = 0.56$, with the cosmological parameters given in Table~\ref{tab:BigMD_param}. The number density of haloes is chosen to reproduce the population of LRGs of the BOSS survey, that have the same typical density \citep[e.g.][]{Alam2017}.

\begin{table}
\caption{Cosmological parameters for the Planck BigMultiDark simulation and \patchy{} mocks, within a flat $\Lambda$CDM framework.}
\centering
\begin{tabular}{cc}
\toprule
Parameter & Value \\
\midrule
$\Omega_{\rm m}$ & 0.307115 \\
$\Omega_{\rm b}$ & 0.048206 \\
$\sigma_8$ & 0.8288 \\
$n_s$ & 0.96 \\
$h$ & 0.6777 \\
\bottomrule
\end{tabular}
\label{tab:BigMD_param}
\end{table}

It has been shown that both two and three-point statistics of the \textsc{BigMD} halo catalogue are accurately reproduced by the \patchy{} mock haloes \citep[][]{Chuang2015}. We then generate 100 realisations of \patchy{} mock halo catalogues using the same input parameters, but varying the random seed for the initial conditions, to estimate the effects of cosmic variance in our analysis.

Furthermore, we have generated another set of 100 \patchy{} realisations, with the same initial conditions and parameters as the realisations above, but replacing the input linear power spectrum with a smoothed power spectrum without BAO wiggles. We dub this set of mocks \patchy{} non-wiggle mocks. Therefore, the only difference between the non-wiggle mocks and the original \patchy{} mocks is the BAO signal.

\subsection{Simulated MultiDark-PATCHY galaxy catalogues}

In addition, to make a robust analysis of uncertainties for the observed data, we use a large number of accurate mock galaxy catalogues.
The MultiDark-\patchy{} \citep[MD-\patchy{},][]{Kitaura2016a} mock galaxy catalogues are constructed based on the mock halo catalogues generated the same way as in \S\ref{sec:patchy_halo_mock}, but with 10 redshift snapshots. The HADRON code \citep[][]{Zhao2015} is then applied to assign mass to haloes, followed by the Halo Abundance Matching \citep[HAM, e.g.][]{Nuza2013} scheme to populate galaxies. Finally, light-cone mocks are built using the SUGAR code \citep[][]{Rodriguez2016}, taking into account observational effects such as stellar mass incompleteness and fibre collisions.

In particular, the HAM model applied to the \patchy{} haloes is calibrated by matching the \textsc{BigMD} simulation with the BOSS DR12 data. Thus, this set of mocks has been named MD-\patchy{} DR12 mocks, as in the BOSS publications. Indeed, they have shown accurate clustering statistics compared to that of the observed data \citep[][]{Kitaura2016a}.
In this work, we use 1000 realisations of MD-\patchy{} galaxy mocks. Furthermore, we divide the galaxy mocks into the same two redshift bins as the case of observed data.

%%%%%%%%%%%%%%%%%%%%%%

\section{Method}
\label{sec:method}

In this section we present the methodology to obtain the 2PCF and its covariance for galaxies and cosmic voids after BAO reconstruction.
%Firstly, we apply the BAO reconstruction technique, which implies tracing galaxies back in time, thereby linearising the peaks of the density field.

\subsection{BAO reconstruction}

The BAO reconstruction technique was introduced by \citet[][]{Weinberg1992} and extended by \citet[][]{Eisenstein2007, Noh2009, Padmanabhan2009, Burden2014}. Since then, it has been widely applied to galaxy catalogues and shown improvements on the precision of BAO peak position measurements in most cases \citep[e.g.][]{Anderson2012, Padmanabhan2012, Anderson2014, Burden2014, Kazin2014, Vargas2015, Alam2017}. Therefore, it has become an essential tool for BAO analyses.

The basic idea of BAO reconstruction is to reverse partially the bulk flow of galaxies. In practice, we smooth the galaxy density field using a Gaussian kernel with a smoothing scale $R$, and then solve the Lagrangian displacement field of galaxies through the Zel'dovich approximation \citep[][]{Zeldovich1970, Croft1997}. With the displacement field we are able to move back galaxies, as well as correct redshift space distortions, thus partially removing the non-linear effects of structure growth.
This procedure has been interpreted as a way to transfer information encoded in the three and four point statistics to the two point correlation function \citep[][]{Schmittfull2015}.

Apart from the `standard' BAO reconstruction scheme, a number of alternative or improved methods have been proposed, such as BAO reconstruction with the `infinity compressible' fluid \citep[][]{Mohayaee2008}, optimal filters \citep[][]{Tassev2012}, effects of local environments \citep[][]{Achitouv2015, Neyrinck2018}, isobaric algorithm \citep[][]{Wang2017}, and iterative algorithm with different smoothing lengths \citep[][]{Schmittfull2017}. Furthermore, some Bayesian methods have been developed to recover accurately the initial density field given a distribution of galaxies \citep[][]{Kitaura2013a, Jasche2013, Wang2013}.

In this work we exploit the standard BAO reconstruction method. In particular, for the \patchy{} cubic halo mocks, we use a smoothing scale of $R = 5\,\hmpc$ for BAO reconstruction. While the reconstructed MD-\patchy{} DR12 galaxy mocks are the same as those used in \citet[][]{Alam2017}, with $R = 15\,\hmpc$. It has been shown that the difference on smoothing scale does not significantly affect the monopole BAO peak position constraint \citep[][]{Vargas2015}. Our studies in this work includes both pre-reconstruction and post-reconstruction cases for all the data and mocks.

\subsection{Void finding}

We apply the \textsc{dive} void finding algorithm \citep[][]{Zhao2016} to the halo or galaxy catalogues to obtain our void samples.
It uses the Delaunay Triangulation \citep[][]{Delaunay1934} technique to partition the volume with a set of discrete points (tracers) into tetrahedra, such that the vertex of the tetrahedra are the tracers, and there is no tracer inside the circumsphere of the tetrahedra.
We dub these empty spheres DT voids, and use the centres of the spheres for clustering analysis.
They have high overlapping fraction as the center of two circumspheres can be very close to each other, resulting in a dense distribution across the cosmic web. Moreover, the centre of circumspheres can be outside the associated tetrahedra.

For the cubic halo mocks, we need to apply periodical boundary conditions. Thus, haloes that are closer than $100\,\hmpc$ to the nearest boundary are copied to the opposite outside of the box. After applying Delaunay Triangulation, we then remove voids found outside the original box. The duplicate length is chosen to be larger than the radius of the largest void ($\sim 40\,\hmpc$). For light-cone mocks, we simply mask out voids centred in unobserved regions.

It is worth noting that the complexity of the \textsc{dive} algorithm scales as $\mathcal{O} (n \log n)$, where $n$ denotes the number of matter tracers (haloes or galaxies). This high efficiency makes our void finding scheme applicable to clustering statistics with thousands of mock realisations for the state of art survey sizes.

Furthermore, \textsc{dive} requires only a halo or galaxy catalogue in co-moving space, without any additional assumptions.
Indeed, each DT void is only described by 4 parameters: 3 for the coordinates (of the centres of circumspheres) and 1 for the radius. In particular, the radii are strongly correlated to the underlying dark matter density field.
It has been shown that the bias and BAO significance vary for different void radius bins, and only large voids are mostly located in under-dense regions of the density field, with anti-correlations to galaxies or haloes \citep[][]{Zhao2016, Chuang2017}. 
This is also consistent with studies in \citet[][]{Schaap2000}, which reveals the ability of probing different density features using DT based methods.
Therefore, a selection of DT voids based on radii is necessary for a clean BAO measurement from under-densities.

In particular, the average radius of all DT voids in the cubic mocks are $\sim 12\,\hmpc$, which is roughly $\sqrt{3}/2$ times of the mean separation of haloes ($\sim 14\,\hmpc$, given the mean number density of $3.5 \times 10^{-4}\,h^3\,{\rm Mpc}^{-3}$).
Indeed, for a uniform halo sample distributed on regular grids, the radii of DT voids are precisely $\sqrt{3}/2$ times the grid size. Thus, it is reasonable that the mean radius of voids is larger than the mean separation.
%For a Poisson random halo catalogue, the expected mean radius of DT voids should be the same. Therefore, the average radius of voids from mocks is consistent with the quasi-Poisson galaxy formation process \citep[e.g.][]{Coles1993}, given the assumption that the density field is homogeneous on large scales. Thus, the radius threshold for voids tracing under-densities should be at least larger than the mean value.

The optimal radius threshold for void selection is defined using mocks, by maximising the signal-to-noise ratio of the BAO peak significance, which is expressed by the difference of the amplitude between the BAO peak and the dips on scales smaller and larger than the peak. In particular, for BOSS DR12 like data, \citet[][]{Liang2016} used the MD-\patchy{} DR12 galaxy mocks and found an optimal radius threshold of $R_{\rm th} = 16\,\hmpc$.

\begin{figure}
\centering
\includegraphics[width=\columnwidth]{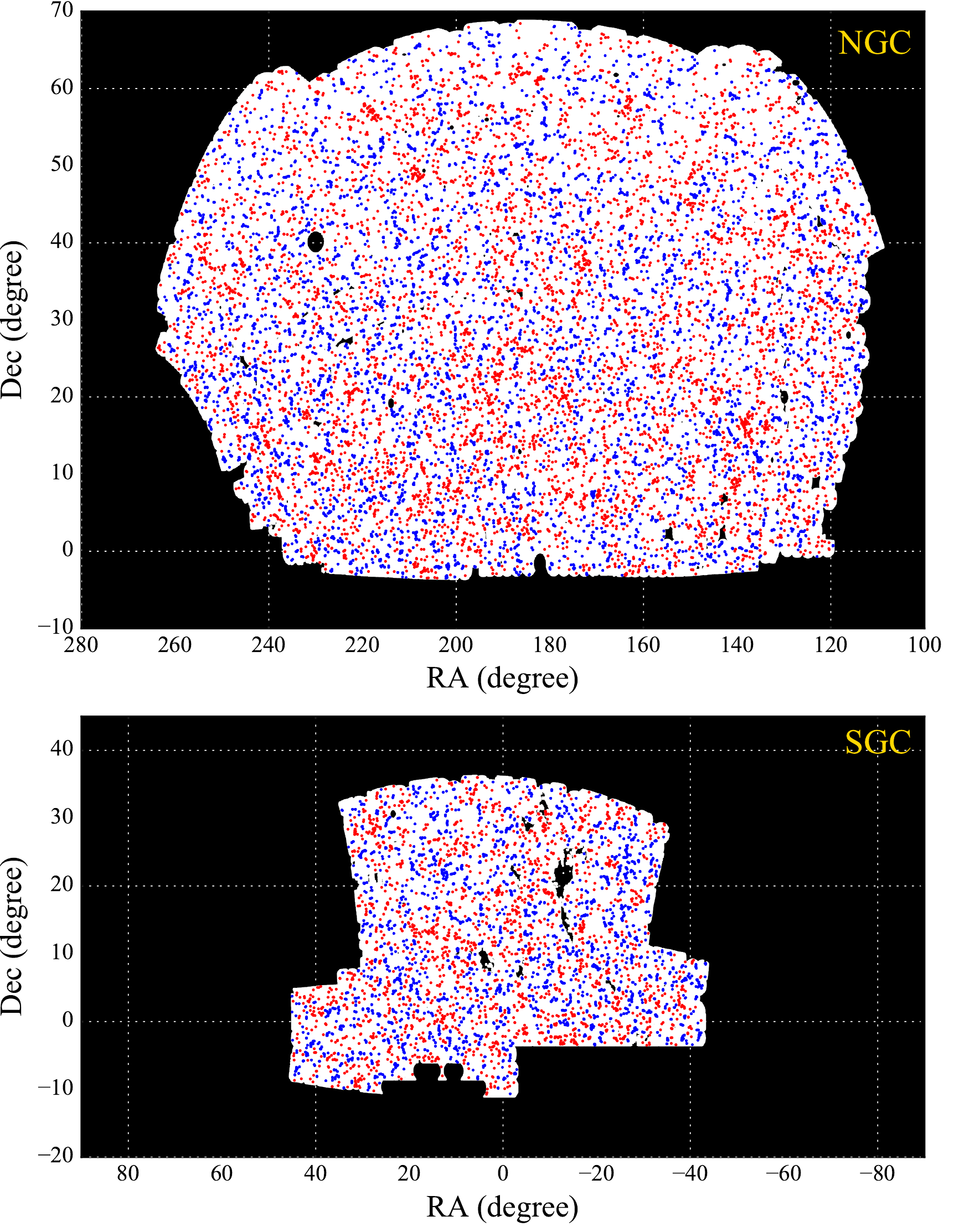}
\caption{Angular distribution of BOSS DR12 LRGs (red points) and the corresponding DT voids with radii $\ge 16\,\hmpc$ (blue points) in the Northern Galactic Cap (NGC, {\it upper} panel) and Southern Galactic Cap (SGC, {\it lower} panel), within a redshift slice of $0.498 < z < 0.5$.}
\label{fig:dr12_sky}
\end{figure}

For simplicity, we use the same selection criteria for the cubic mocks ($R_{\rm V} \ge 16\,\hmpc$), as they have roughly the same number density as well as clustering power as the MD-\patchy{} DR12 mocks. In this case, $\sim 75\,\%$ of the voids are removed, but the number of remaining voids is still $\sim 1.5$ times as that of haloes/galaxies. Furthermore, it has been confirmed in \citet[][]{Zhao2016} that over 80\,\% of the voids with radii $R_{\rm V} \ge R_{\rm th}$ reside in dynamical expanding regions.
A slice of the distribution of galaxies and voids from BOSS DR12 data is shown in Figure~\ref{fig:dr12_sky}, in which voids appear to fill the gaps between galaxies.

Since voids are indirect observables, and are defined by haloes or galaxies, we cannot directly apply BAO reconstruction to voids by reversing the displacement of density troughs. Instead, we run BAO reconstruction for haloes or galaxies first, followed by \textsc{dive} to construct the post-reconstruction void sample. We note that this removes constraining power from voids and transfers it to haloes. For the light-cone mocks, we still use the original survey masks to remove voids outside the observed area.

In this work, we consider neither systematical weights nor FKP weights for voids. Nevertheless, comparing the 2PCFs of voids to mocks, we do not see significant differences \citep[][]{Kitaura2016, Liang2016}. We leave the detailed weighting scheme for voids to a future work.

\subsection{Correlation function estimator}
Two-point correlation function describes the excess probability of finding a galaxy pair compared to a random distribution. Therefore, the pair counts from both the data catalogue and the random distribution are necessary for estimating 2PCFs. To obtain unbiased correlation functions with the minimal statistical noise introduced by a finite random sample, a number of correlation function estimators are proposed, such as the Peebles--Hauser estimator \citep[][]{Peebles1974}:
\begin{equation}
\xi_{\rm box} (s) = \frac{DD (s)}{RR (s)} - 1 ,
\end{equation}
where $DD(s)$ and $RR (s)$ are the number of tracer pairs with separation $s$ from the data-data and random-random catalogues, normalised by the corresponding number of total pairs, respectively.

Fortunately, for a periodical cubic box, the pair counts of a uniform random sample can be calculated analytically, i.e.,
\begin{equation}
RR (s) = \frac{4 \uppi}{3} \frac{s_{\rm max}^3 - s_{\rm min}^3}{2 V} ,
\end{equation}
where $s_{\rm max}$ and $s_{\rm min}$ are the upper and lower boundaries of the separation bins, and $V$ denotes the co-moving volume of the data catalogue. Therefore, for mock catalogues in cubic boxes, we use simply the Peebles--Hauser estimator together with the analytical $RR$ term.

We compute the monopole 2PCF $\xi (s)$ for the pre-reconstruction light-cone samples (including MD-\patchy{} mocks and BOSS data) using the Landy--Szalay estimator \citep[][]{Landy1993}, since it is in general a better choice than the Peebles--Hauser estimator with a finite random catalogue \citep[e.g.][]{Pons1999, Kerscher2000, Vargas2013}:
\begin{equation}
\xi_{\rm pre} (s) = \frac{DD (s) - 2 DR (s) + RR (s)}{RR (s)},
\label{eq:LS}
\end{equation}
where $DR (s)$ indicates the number of data-random pairs at each separation bin, normalised by the total number of pairs.
We use a uniformly distributed random catalogue for the cubic mocks, which contains the same number of tracers as the data, and random catalogues generated using the `shuffled' method \citep[][]{Ross2012, Liang2016} for light-cone mocks.

For the post-reconstruction light-cone samples, besides reversing the displacement of haloes or galaxies, we also shift the random catalogue by the same displacement field. 
Denoting the shifted random catalogue by $S$, the Landy--Szalay estimator becomes \citep[][]{Padmanabhan2012}
\begin{equation}
\xi_{\rm post} (s) = \frac{DD (s) - 2 DS (s) + SS (s)}{RR (s)} ,
\label{eq:cfpost}
\end{equation}
where $D$ is the shifted data catalogue, and $R$ is the same random catalogue as the pre-reconstruction case.

Since voids are reconstructed by applying the void finder to post-reconstruction halo/galaxy catalogues, there is no displacement field for shifting the randoms. Therefore, we have to use Eq.~\ref{eq:LS} for estimating the post-reconstruction correlation functions for voids, which does not require the $S$ catalogue. We construct $R$ catalogues for pre- and post-reconstruction catalogues separately using the shuffled method described in \citet[][]{Liang2016}.

%we do not have a specific displacement field for voids, we are not able to construct the corresponding $S$ catalogue. Therefore, we use still the shuffled method to generate the random void catalogue for the post-reconstruction case. Consequently, Eq.~\ref{eq:LS} is also used for reconstructed voids.

Throughout this paper, the range of separation for the 2PCFs is always $s \in [0, 200]\,\hmpc$, with the bin size of $5\,\hmpc$.
Furthermore, the random catalogues for cubic mocks contains the same number of tracers as the data, and the number of (shifted) random tracers for the light-cone mocks are always 20 times that of the corresponding data for both galaxies and voids.

%The centre of the bins are computed by
%\begin{equation}
%s_c = ( s_{\rm min} + s_{\rm max} ) / 2,
%\end{equation}
%where $s_{\rm min}$ and $s_{\rm max}$ are the lower and upper boundaries of the bins.

\subsection{Covariance matrices}

We estimate the covariance matrices of the 2PCFs by using the sample covariance of mocks:
\begin{equation}
\textbf{C}_{s, ij} = \frac{1}{N_{\rm m} - 1} \sum_{k=1}^{N_{\rm m}} [ \xi_k (s_i) - \bar{\xi} (s_i) ][ \xi_k (s_j) - \bar{\xi} (s_j) ],
\end{equation}
where $N_{\rm m}$ is the number of mocks, $\xi_k (s)$ is the 2PCF of the $k$-th mock, and $\bar{\xi} (r)$ is the mean 2PCF of all the mocks.

Then, to evaluate the $\chi^2$ for the fit, we estimate the unbiased inverse covariance matrix as \citep[][]{Hartlap2007, Percival2014}
\begin{equation}
\textbf{C}^{-1} = \textbf{C}_s^{-1} \frac{N_{\rm m} - N_{\rm bins} - 2}{N_{\rm m} - 1},
\end{equation}
where $N_{\rm bins}$ is the number of data bins used for the fit.

%%%%%%%%%%%%%%%%%%%%%%

\section{BAO fitting}
\label{sec:fit}

The typical de-wiggled model for galaxies \citep[][]{Xu2012} does not work well for voids (cf. \S\ref{sec:res1}), mainly due to the oscillation patterns induced by void exclusion effects.
In fact, voids show radii-dependent negative correlations on small scales \citep[Figures 1 and 2 in][]{Liang2016}. This feature is consistent with the oscillation patterns of the void power spectra on large $k$, which do not affect the BAO signature \citep[][]{Chan2014, Hamaus2014b, Zhao2016}.
Therefore, we present in this section an improved BAO model, and discuss some details for the fits.
In particular, the fitting range we choose in this section is $s \in [60, 160]\,\hmpc$, which is large enough to cover the BAO peak feature in the 2PCFs.

\subsection{The typical de-wiggled model}

We follow \citet[][]{Xu2012} for computing the template 2PCF:
\begin{equation}
\xi_{\rm t} (s) = \int \frac{k^2 \,{\rm d}k}{2 \uppi^2} P_{\rm t} (k) j_0 (k s) \, {\rm e}^{-k^2 a^2} ,
\label{eq:xi_model}
\end{equation}
where $P_{\rm t} (k)$ is the template power spectrum, $j_0$ is the 0-order spherical Bessel function of the first kind (the sinc function), and $a$ is a factor for the Gaussian high-$k$ damping. We use $a = 1\,\hmpc$, as in \citet[][]{Xu2012}.

The template power spectrum is generated by
\begin{equation}
P_{\rm t} (k) = [ P_{\rm lin} (k) - P_{\rm lin, nw} (k) ] \, {\rm e}^{-k^2 \Sigma_{\rm nl}^2 / 2} + P_{\rm lin, nw} (k).
\label{eq:pm}
\end{equation}
Here, $P_{\rm lin} (k)$ is the linear power spectrum, $P_{\rm lin, nw} (k)$ is the non-wiggle (free of BAO wiggles) power spectrum, and $\Sigma_{\rm nl}$ is the damping parameter for BAO.
The first term on the right hand side indicates the BAO wiggles, while the second term describes the broad-band feature. Thus, $P_{\rm t} (k)$ is called the `de-wiggled' power spectrum.

For the cubic halo mocks, we use the input power spectra when generating the mocks for the fitting, hence the fiducial cosmology is exactly the `true' cosmology. For the light-cone mocks and BOSS DR12 data, $P_{\rm lin} (k)$ is obtained by the CAMB\footnote{\url{http://camb.info}} \citep[][]{Lewis1999} software, and $P_{\rm lin, nw} (k)$ is computed using the fitting formulae of \citet[][]{Eisenstein1998}, with the fiducial cosmological parameters given by Table~\ref{tab:fiducial}, to be consistent with the BOSS publications.

\begin{table}
\caption{The fiducial cosmological parameters for the co-moving space sample construction and BAO fitting, within a flat $\Lambda$CDM framework.}
\centering
\begin{tabular}{cc}
\toprule
Parameter & Value \\
\midrule
$\Omega_{\rm m}$ & 0.31 \\
$\Omega_{\rm b}$ & 0.048143 \\
$\sigma_8$ & 0.8 \\
$n_s$ & 0.97 \\
$h$ & 0.676 \\
\bottomrule
\end{tabular}
\label{tab:fiducial}
\end{table}

Then, we fit the 2PCFs using the forms in \citet[][]{Xu2012}:
\begin{equation}
\xi_{\rm model} (s) = B^2 \xi_{\rm t} (\alpha s) + A(s) ,
\label{eq:xim}
\end{equation}
where
\begin{equation}
A(s) = \frac{a_1}{s^2} + \frac{a_2}{s} + a_3.
\label{eq:nuisance}
\end{equation}
Here, $B$ is the normalisation parameter (it absorbs the bias factor $b$), $\alpha$ is the scale dilation parameter, and $a_1$, $a_2$, and $a_3$ are the linear nuisance parameters.

Indeed, $\alpha$ is the measurement of the baryon acoustic scale (the BAO peak position) relative to the fiducial cosmology, which can be expressed as
\begin{equation}
\alpha = \alpha_\perp^{2/3} \alpha_\parallel^{1/3},
\end{equation}
where $\alpha_\perp$ and $\alpha_\parallel$ are the dilation in the transverse and line-of-sight directions respectively:
\begin{subequations}
\begin{align}
\alpha_\perp &= \frac{D_{\rm A} (z) r_{\rm d}^{\rm fid}}{D_{\rm A}^{\rm fid} (z) r_{\rm d}} \\
\alpha_\parallel &= \frac{H^{\rm fid} (z) r_{\rm d}^{\rm fid}}{H (z) r_{\rm d}} .
\end{align}
\end{subequations}
Here, $D_{\rm A} (z)$ and $H(z)$ are the angular diameter distance and Hubble parameter respectively, and $r_{\rm d}$ is the sound horizon at radiation drag.

In summary, we have 6 free parameters in this model, they are $\Sigma_{\rm nl}, B, a_1, a_2, a_3$, and $\alpha$.

\subsection{Parameter inference}

Since the nuisance parameters $a_1$, $a_2$, and $a_3$ are for the modelling of the broad-band shape of the correlation function, including scale-dependent bias, non-linear evolution, and observational systematic effects, they have minor impacts on the fitting of the BAO peak position \citep{Xu2012, Ross2017}. Therefore, to simplify and accelerate the parameter inference procedure, we use a linear least-square estimator to obtain the best-fit value for these 3 parameters.

For the rest of the parameters we perform a Monte-Carlo Bayesian posterior sampling, to explore the likelihood distribution in the parameter space. In particular, we assume a Gaussian likelihood:
\begin{equation}
\mathcal{L} \equiv p({\rm data} | \boldsymbol{\Theta}, {\rm model}) \propto \exp{(- \chi^2 (\boldsymbol{\Theta}) / 2)} ,
\end{equation}
where $\boldsymbol{\Theta}$ indicates the set of fitting parameters, and $p({\rm data} | \boldsymbol{\Theta}, {\rm model})$ is the joint probability distribution of the observed 2PCF, given the model and the parameters. Moreover, $\chi^2$ can be obtained by
\begin{equation}
\chi^2 (\boldsymbol{\Theta}) = ( \boldsymbol{\xi}_{\rm data} - \boldsymbol{\xi}_{\rm model} (\boldsymbol{\Theta}) )^T \textbf{C}^{-1} ( \boldsymbol{\xi}_{\rm data} - \boldsymbol{\xi}_{\rm model} (\boldsymbol{\Theta}) ) .
\end{equation}
Here, $\boldsymbol{\xi}_{\rm data}$ indicates the measured 2PCF, and $\boldsymbol{\xi}_{\rm model}$ is the 2PCF evaluated from the model.

The Bayesian posterior distribution is then
\begin{equation}
p (\boldsymbol{\Theta} | {\rm data}, {\rm model}) = \frac{p({\rm data} | \boldsymbol{\Theta}, {\rm model}) p(\boldsymbol{\Theta} | {\rm model})}{p({\rm data} | {\rm model})} ,
\end{equation}
where $p(\boldsymbol{\Theta} | {\rm model})$ is the prior distribution for the parameters, and $p({\rm data} | {\rm model})$ is the Bayesian evidence $\mathcal{Z}$, given by
\begin{equation}
\mathcal{Z} \equiv p({\rm data} | {\rm model}) = \int p({\rm data} | \boldsymbol{\Theta}, {\rm model}) p(\boldsymbol{\Theta} | {\rm model}) \, {\rm d} \boldsymbol{\Theta} ,
\end{equation}
which is a good indicator for model selection \citep[e.g.][]{Mukherjee2006}.

In this work, we choose flat priors for the $\alpha$, $B$, and $\Sigma_{\rm nl}$ parameters in the following ranges
\begin{subequations}
\begin{align}
\alpha &\in [0.8, 1.2] ,\\
B &\in [0, 50] ,\\
\Sigma_{\rm nl} &\in [0, 25] \,\hmpc .
\end{align}
\end{subequations}
We shall see from the fit results in the next sections that the prior ranges are large enough compared to the posterior of the parameters.

In practice, we rely on the \textsc{MultiNest}\footnote{\url{https://ccpforge.cse.rl.ac.uk/gf/project/multinest/}} \citep[][]{Feroz2008, Feroz2009, Feroz2013} tool for accurate and efficient parameter estimation and model selection. It uses the Nested Sampling method \citep{Skilling2004} to calculate the Bayesian evidence, and provides also the posterior inferences of parameters.
Furthermore, we marginalise the posterior distributions and obtain the $1\,\sigma$ confidence intervals using the \textsc{GetDist}\footnote{\url{https://github.com/cmbant/getdist}} package. The best-fit value of parameters given in this paper are simply the mean of the lower and upper $1\,\sigma$ confidence limits.

\subsection{Fit results with the typical de-wiggled model}
\label{sec:res1}

\begin{figure}
\centering
\includegraphics[width=\columnwidth]{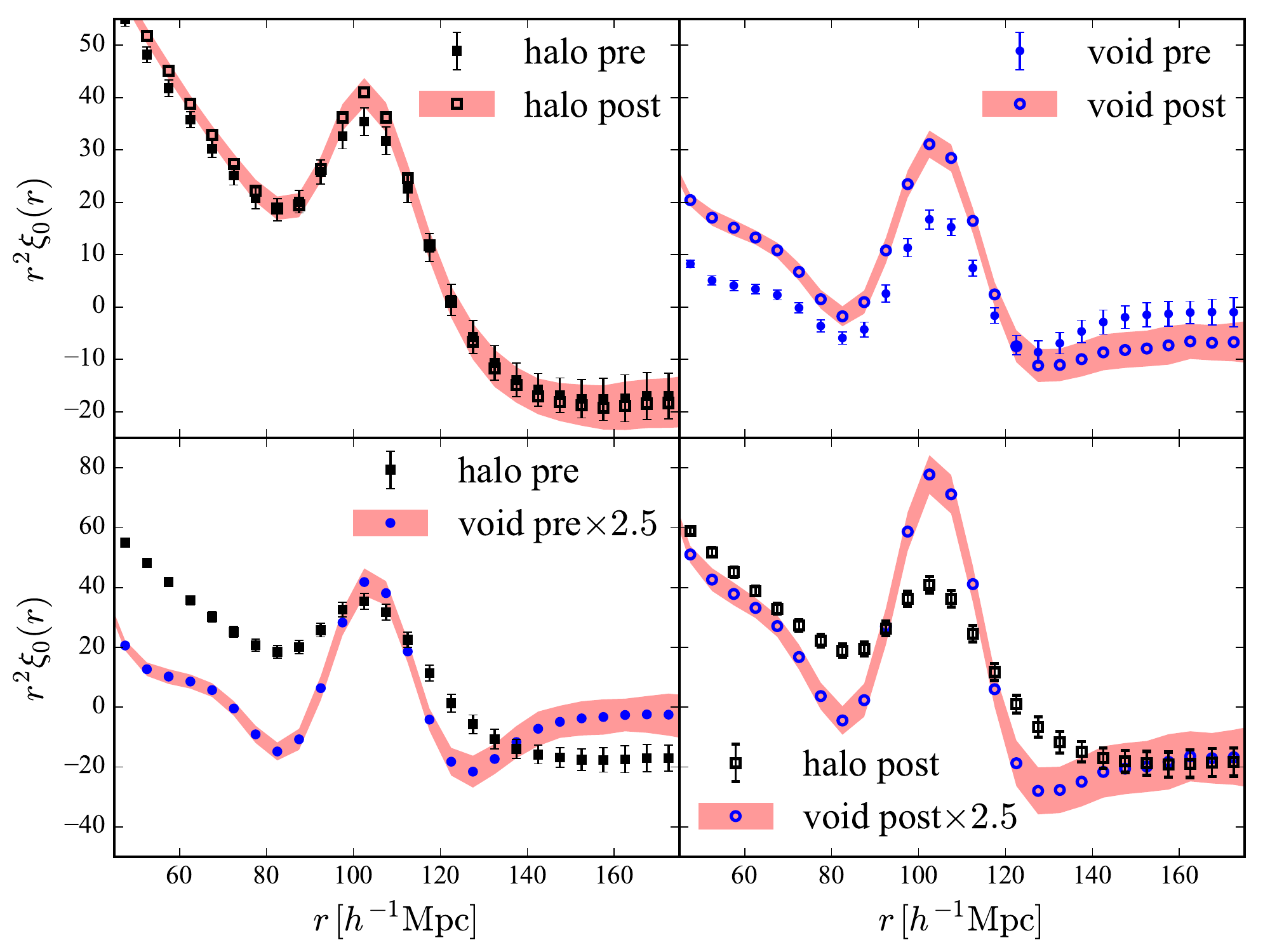}
\caption{The comparisons of 2PCFs of haloes and voids from \patchy{} cubic halo mocks. `pre' in the legends indicates results from the samples without BAO reconstruction, while `post' denotes results with BAO reconstruction. The error bars and shadowed bands show errors obtained from 100 realisations. For the comparisons between haloes and voids, we multiply the 2PCF of voids by a factor of 2.5 to obtain similar amplitudes on BAO scale.}
\label{fig:halo_cf}
\end{figure}

To verify our BAO fitting method, we apply the fitting procedure to the 2PCFs of both pre-reconstruction haloes and voids from the \textsc{Patchy} cubic halo mocks. The comparison of 2PCFs of haloes and voids are shown in Figure~\ref{fig:halo_cf}. To reduce the impact of cosmic variance, we fit the model to the mean 2PCF of 100 realisations of mocks, with the covariance matrix drawn from the same suites of mocks.
In this case, the error on the 2PCF to be fitted is overestimated by a factor of $\sim 10$.

Furthermore, using the wiggle free realisations of \textsc{Patchy} mocks, we split the 2PCFs as well as the best-fit model curves into BAO (wiggle) and continuous (non-wiggle) components, and then decompose the model to fit the two parts respectively to check the robustness of the model.

\begin{figure}
    \centering
    \includegraphics[width=\columnwidth]{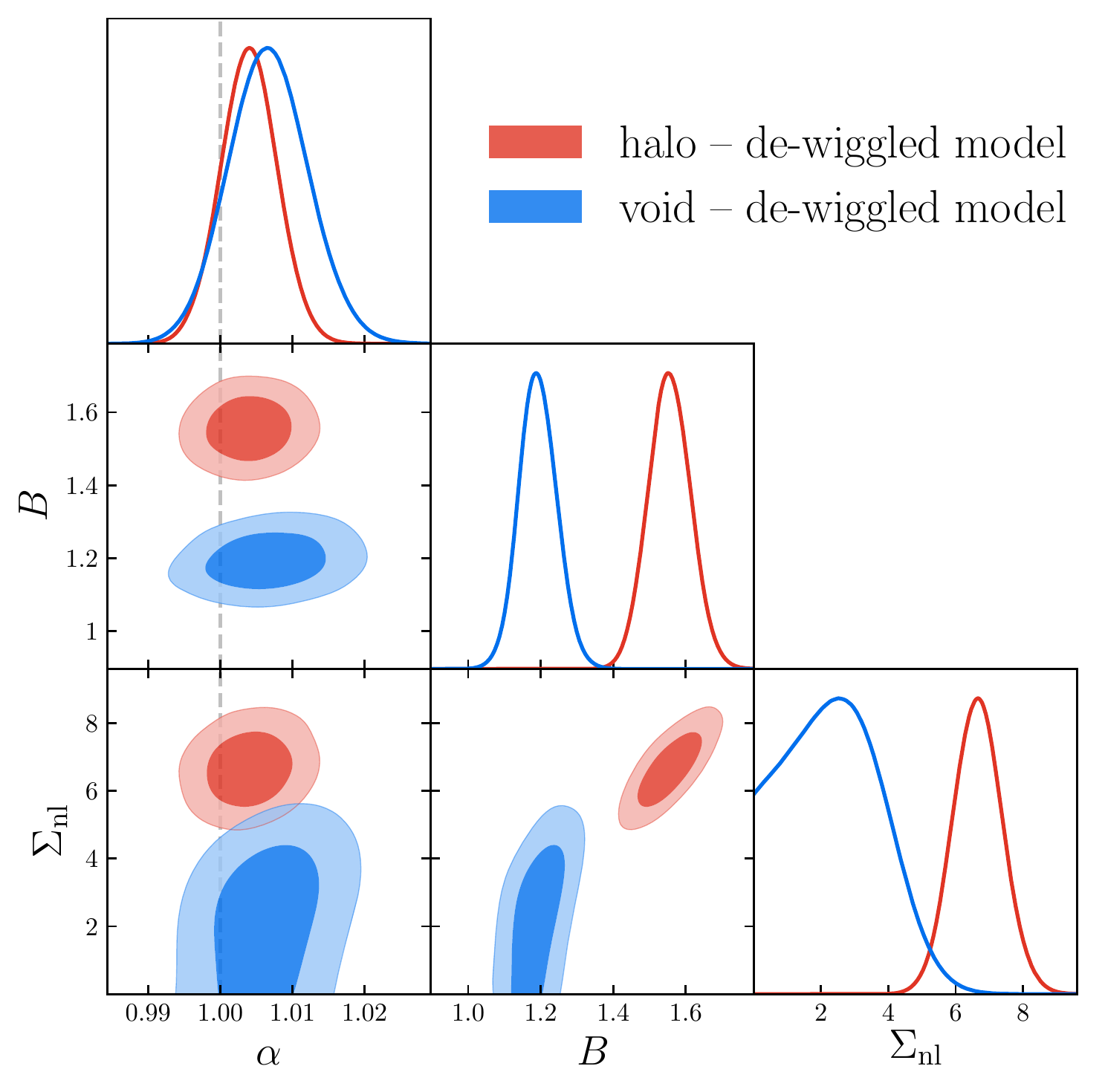}
    \caption{The posterior distribution of parameters of the de-wiggled model, fitted to the mean 2PCF of 100 pre-reconstruction \patchy{} cubic mock {\it halo} (red) and {\it void} (blue) catalogues. The grey dashed line indicates $\alpha = 1$, which is expected $\alpha$ value given the fiducial cosmology.}
    \label{fig:triangle_box_orig_model}
\end{figure}

\begin{figure*}
    \centering
    \includegraphics[width=.9\columnwidth]{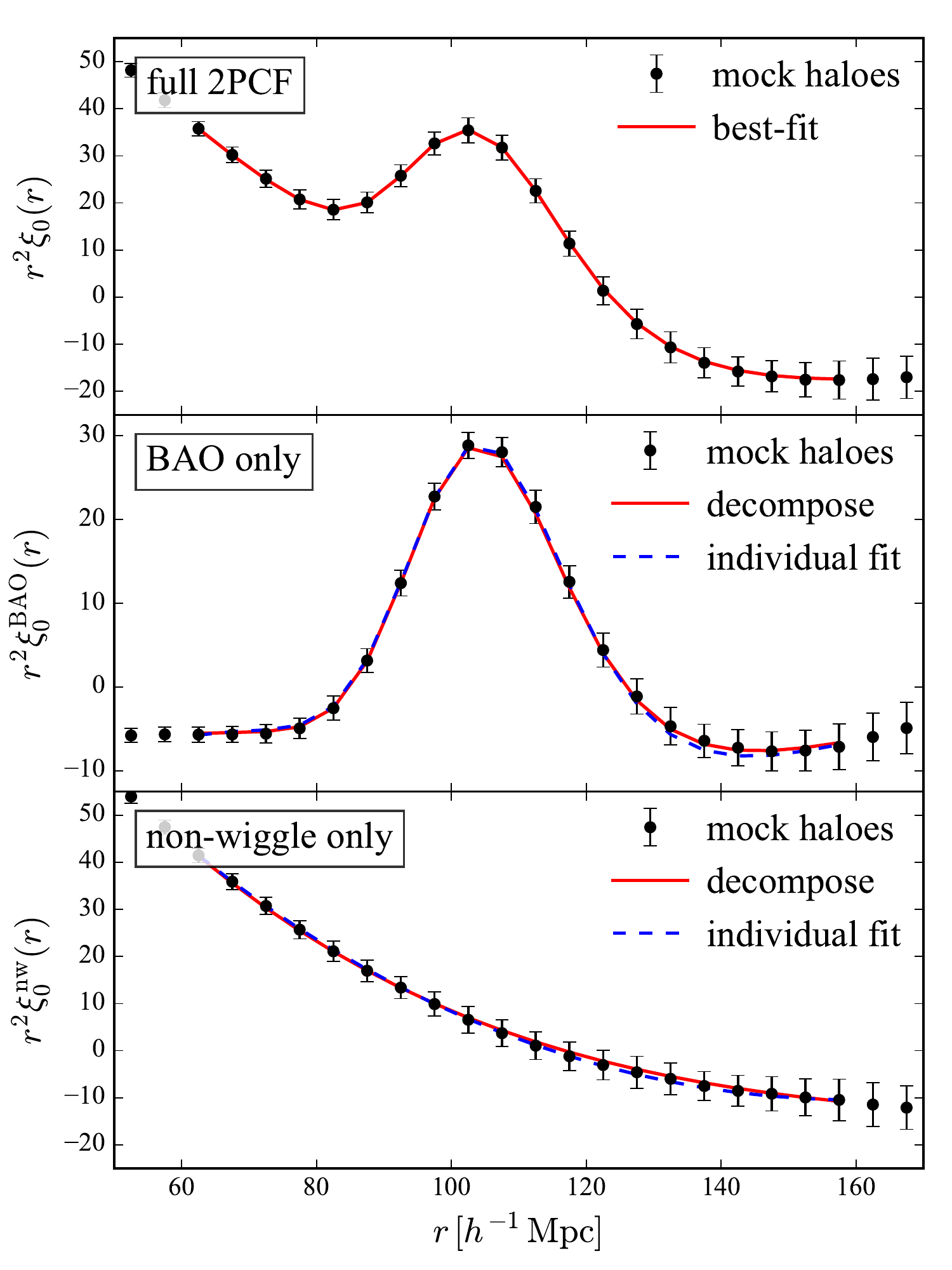}
    \includegraphics[width=.9\columnwidth]{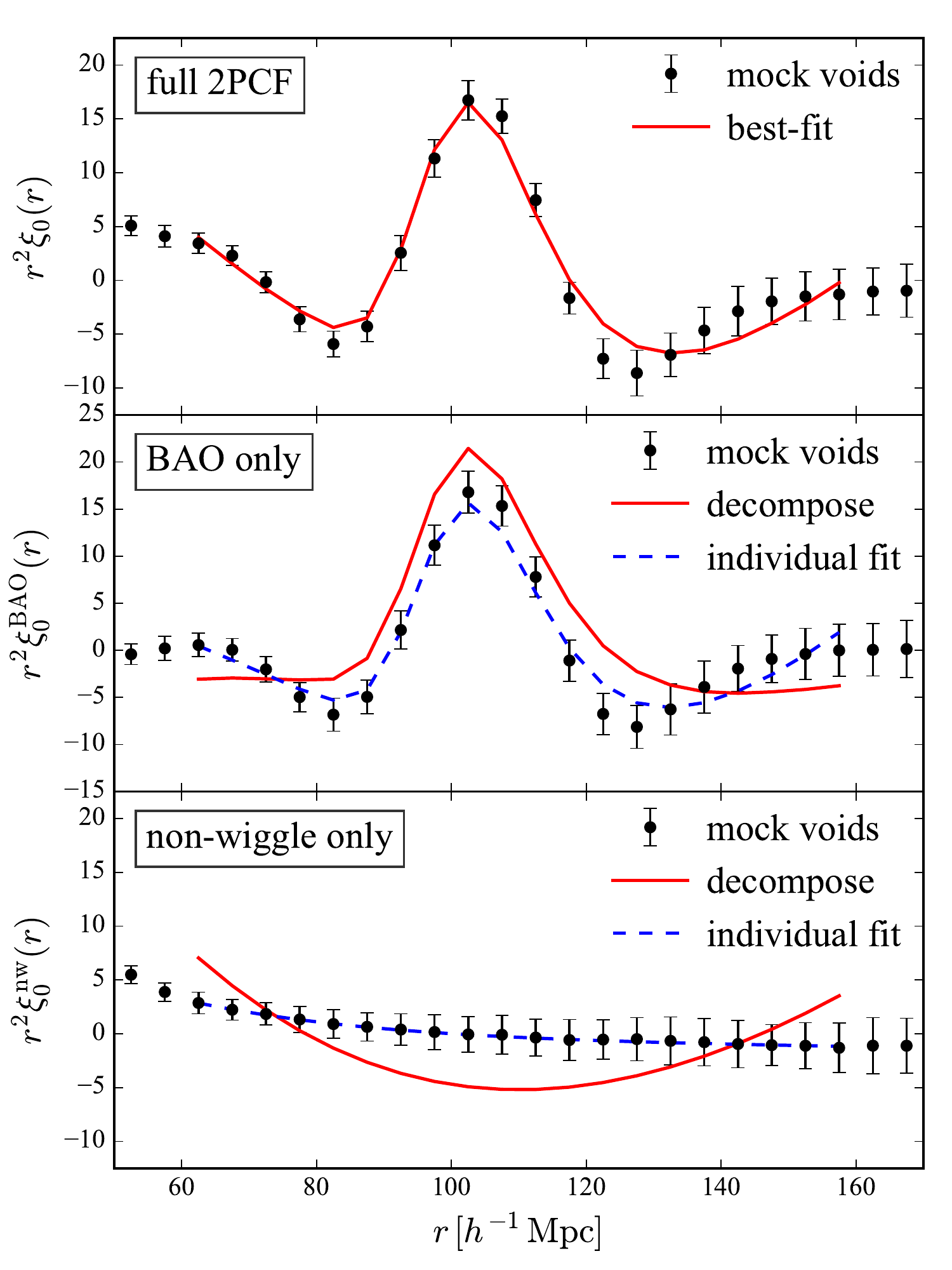}
    \caption{The mean and error of the 2PCFs from 100 \textsc{Patchy} mock {\it halo} ({\it left}) and {\it void} ({\it right}) catalogues (black dots) as well as the best-fit model curves (lines). The red solid lines in the {\it middle} and {\it bottom} panels are directly decomposed from the best-fit curve of the full 2PCF (red line in the first panel), while the blue dashed lines for the last two panels are the best-fit results of the decomposed models.}
    \label{fig:bestfit_box_orig_model}
\end{figure*}

% \begin{figure}
% \centering
% \includegraphics[width=\columnwidth]{fig/void_orig_model}
% \caption{The mean and error of the 2PCFs from 100 \textsc{Patchy} mock {\it void} catalogues (black dots) as well as the best-fit model curves. The blue dashed lines in the second and third panels are directly decomposed from the best-fit curve of the full 2PCF (red line in the first panel), while the red solid lines for the last two panels are the best-fit results of the decomposed models.}
% \label{fig:void_orig_model}
% \end{figure}

The posterior distribution of the parameters, as well as the best-fit (maximum likelihood) curves, are shown in Figure~\ref{fig:triangle_box_orig_model} and Figure~\ref{fig:bestfit_box_orig_model}, respectively.
One can see that the typical de-wiggled model works well for haloes. However, the posterior of $\Sigma_{\rm nl}$ for voids is asymmetric. Moreover, the BAO component of the best-fit model is sub-optimal for voids.
We further list the best-fit value and $1\,\sigma$ error of the fitting parameters, as well as the Bayesian evidence and minimum $\chi^2$ value in Table~\ref{tab:fit_orig_model}.
The Bayesian evidence for voids is much smaller than that of haloes, and it implies that the de-wiggled model is disfavoured by the measured void 2PCF from \patchy{} mocks. The large minimum $\chi^2$ value also indicates that the model does not fit well the void 2PCF.

%The results for haloes and voids are shown in Figure~\ref{fig:halo_orig_model} and Figure~\ref{fig:void_orig_model} respectively. One can see that the typical de-wiggled model works well for haloes, but the wiggled component of the model is sub-optimal for voids. We further list explicitly the fitted parameters in Table~\ref{tab:fit_orig_model}. The large minimum $\chi^2$ value also suggests that the BAO model can be improved.

\begin{table}
\caption{
Fit results of the de-wiggled model for the mean 2PCF of 100 pre-reconstruction \textsc{Patchy} cubic mocks. The confidence intervals are drawn from the posterior shown in Figure~\ref{fig:triangle_box_orig_model}, and $\mathcal{Z}$ indicates the Bayesian evidence.
}
\centering
\begin{tabular}{ccc}
\toprule
Parameter & halo & void \\
\midrule
$\alpha - 1$ & $0.00398 \pm 0.00386$ & $0.00664 \pm 0.00546$ \\
$B$ & $1.56 \pm 0.058$ & $1.19 \pm 0.051$ \\
$\Sigma_{\rm nl}$ & $6.65 \pm 0.726$ & $2.18 \pm 1.482$ \\
$\ln{\mathcal{Z}}$ & $-10.8 \pm 0.10$ & $-16.5 \pm 0.16$ \\
$\chi_{\rm min}^2$ & 0.196 & 14.2 \\
%$\bar{\alpha} - 1$ & 0.00394 & 0.00534 \\
%$\sigma_\alpha$ & 0.00389 & 0.00528 \\
%$\Sigma_{\rm nl}$ & 6.595 & 0 \\
%$B$ & 1.555 & 1.169 \\
%$\chi_{\rm min}^2$ & 0.198 & 14.9 \\
\bottomrule
\end{tabular}
\label{tab:fit_orig_model}
\end{table}

\subsection{The modified de-wiggled model}
\label{sec:bao_model}
The bottom right panel of Figure~\ref{fig:bestfit_box_orig_model} shows that
the broad-band (non-wiggle) terms of the de-wiggled model is accurate for the continuous component of the measured void 2PCF.
We then focus on improving the wiggled component of the model, which comes from the first term on the right hand side of Eq.~\ref{eq:pm}:
\begin{equation}
P_{\rm t}^{\rm w} (k) = [ P_{\rm lin} (k) - P_{\rm lin,nw} (k) ] \, {\rm e}^{-k^2 \Sigma_{\rm nl}^2 / 2} .
\end{equation}

\begin{figure}
    \centering
    \includegraphics[width=.9\columnwidth]{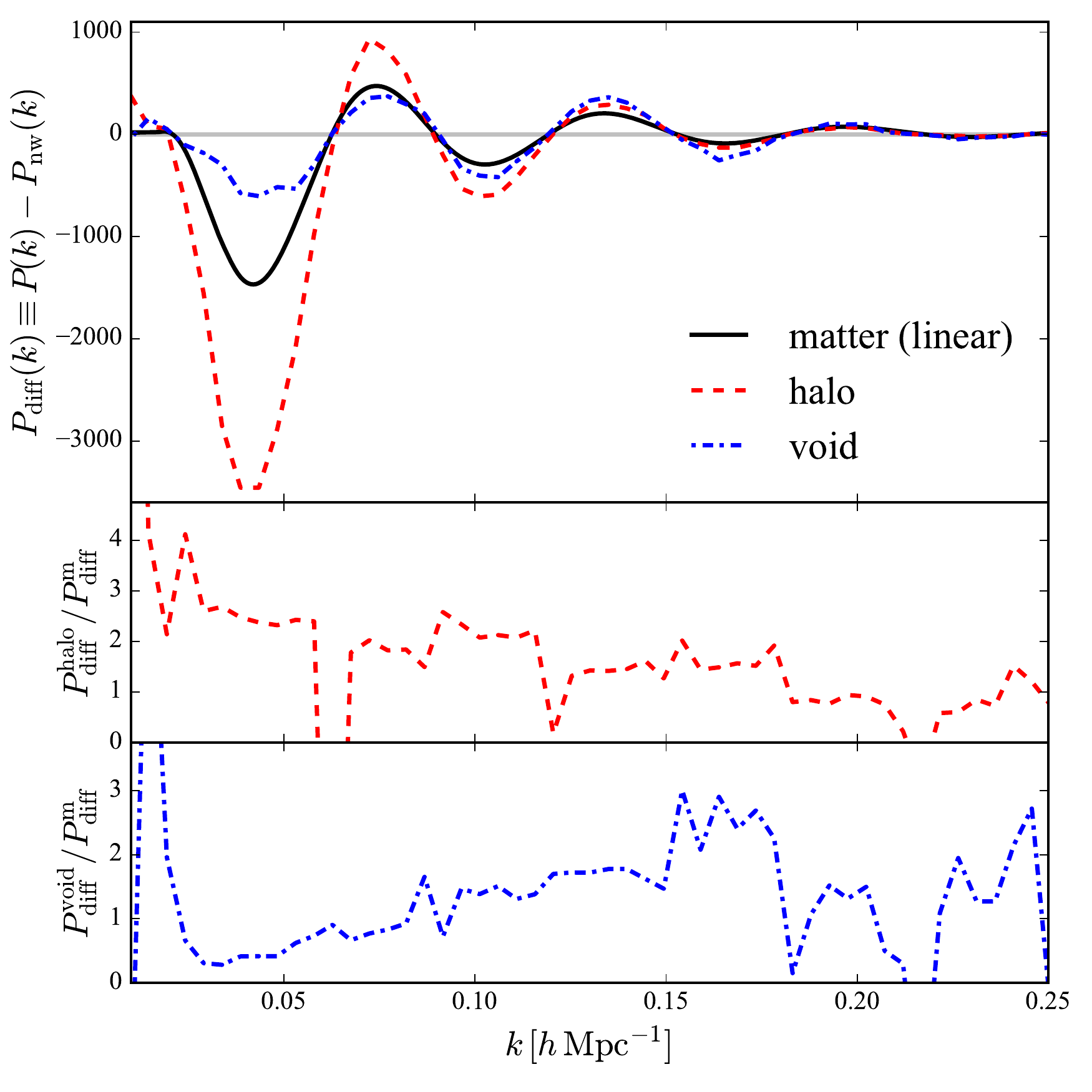}
    \caption{The wiggled components of the power spectra (computed by the difference of the normal power spectra and the non-wiggle ones) for dark matter in the linear regime, haloes, and voids (upper panel), together with the ratio of halo and void BAO to that of dark matter (middle and bottom panels). The increasing ratio in the bottom panel indicated that the wiggles of void clustering cannot be modelled by a damping of the linear wiggles.}
    \label{fig:bao_diff}
\end{figure}

The BAO signal here is modelled by the linear and wiggle-free matter power spectra, we use the damping factor ${\rm e}^{-k^2 \Sigma_{\rm nl}^2 / 2}$ and a bias factor $B$ (Eq.~\ref{eq:xim}) to approximate the halo BAO feature. However, this estimation is not guaranteed to be valid for voids as well. To verify this, using the non-wiggle \textsc{Patchy} mocks, we plot the difference between the power spectrum and the corresponding non-wiggle one for both galaxies and voids, i.e.
\begin{subequations}
\begin{align}
P_{\rm diff}^{\rm halo, w} (k) =  P^{\rm halo} (k) - P_{\rm nw}^{\rm halo} (k) , \\
P_{\rm diff}^{\rm void, w} (k) =  P^{\rm void} (k) - P_{\rm nw}^{\rm void} (k) ,
\end{align}
\end{subequations}
and then compare them with that of dark matter in the linear regime in Figure~\ref{fig:bao_diff}.
In particular, the middle and bottom panels show the damping factor of haloes and voids respectively. The exponential form of the damping factor requires a descending trend with respect to $k$. And this is consistent with the behaviour for haloes. However, the damping factor of voids shows an increasing trend in the range of $k \in [0.05, 0.15]\,h\,{\rm Mpc}^{-1}$. Therefore, the de-wiggled BAO model is suboptimal for voids, and needs to be revised.

\begin{figure}
    \centering
    \includegraphics[width=.9\columnwidth]{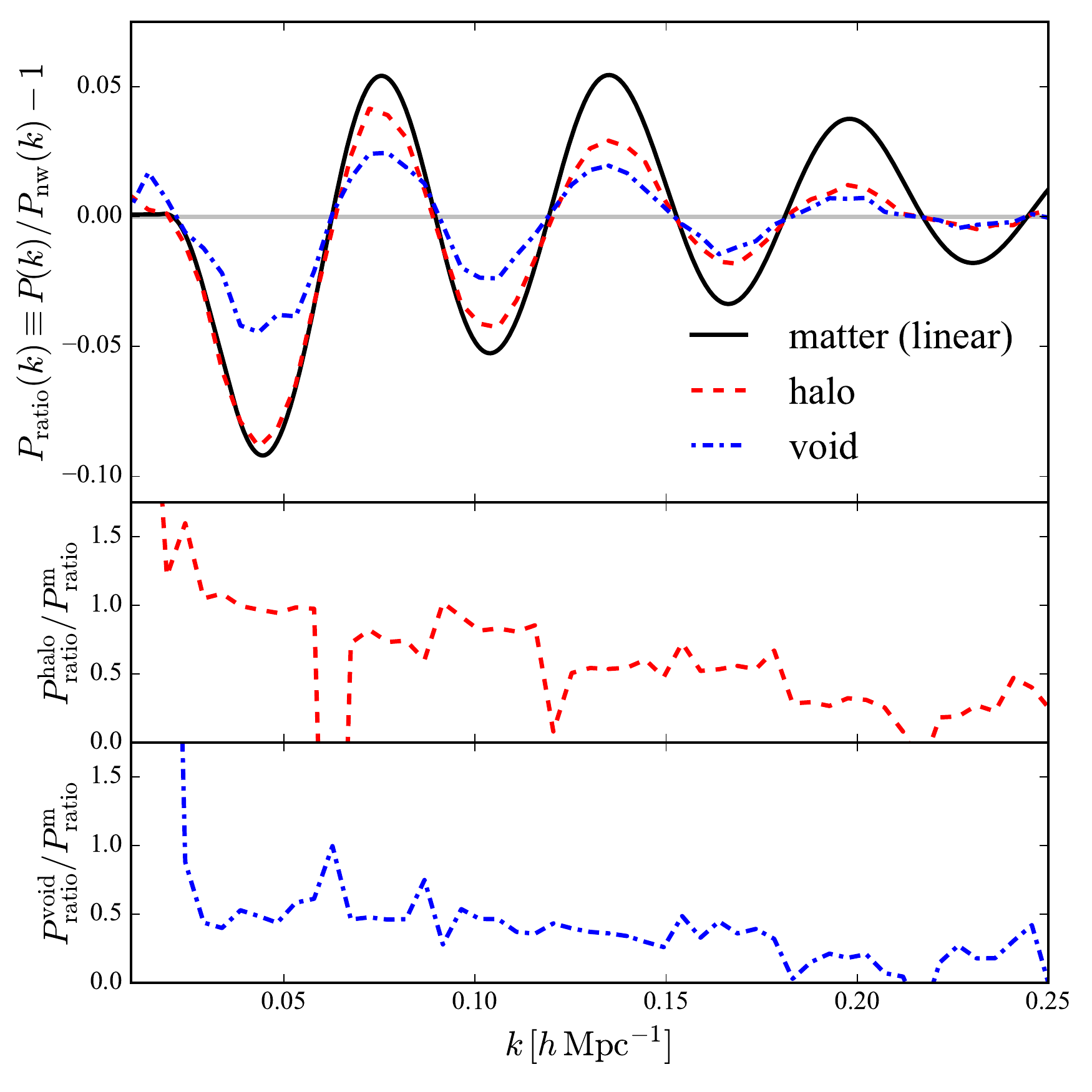}
    \caption{Same as Figure~\ref{fig:bao_diff}, but the wiggled components are computed by the ratio of the normal power spectra to the non-wiggle ones. It is possible to model the wiggles of voids using a damping of the linear wiggles in this case.}
    \label{fig:bao_ratio}
\end{figure}

We then use the ratio of the power spectrum and the corresponding non-wiggle one to indicate the BAO signal, i.e.
\begin{equation}
P^{\rm w \prime} (k) = \frac{ P (k) }{ P_{\rm nw} (k)} - 1,
\end{equation}
and plot the results in Figure~\ref{fig:bao_ratio}. It shows that for both haloes and voids, the wiggled components expressed in this way can be estimated by the linear one with a damping factor. Accordingly, we construct a model that is similar but different to that in Eq.~\ref{eq:pm}:
\begin{equation}
\frac{ P_{\rm t} (k) }{ P_{\rm t, \rm nw} (k) } - 1 =
\left( \frac{ P_{\rm lin} (k) }{ P_{\rm lin, nw} (k) } - 1 \right) \, {\rm e}^{- k^2 \Sigma_{\rm nl}^{\prime 2} / 2} ,
\end{equation}
namely
\begin{equation}
\begin{aligned}
&P_{\rm t} (k) = \\
&\quad \left[ ( P_{\rm lin} (k) - P_{\rm lin, nw} (k) ) {\rm e}^{-k^2 \Sigma_{\rm nl}^{\prime 2} / 2} + P_{\rm lin, nw} (k) \right] \cdot \frac{ P_{\rm t, nw} (k) }{ P_{\rm lin, nw} (k) } .
\end{aligned}
\label{eq:pm1}
\end{equation}
Note that compared to Eq.~\ref{eq:pm}, this new power spectra model contains  only an extra factor $P_{\rm t, nw} (k) / P_{\rm lin, nw} (k)$.

\begin{figure}
    \centering
    \includegraphics[width=.9\columnwidth]{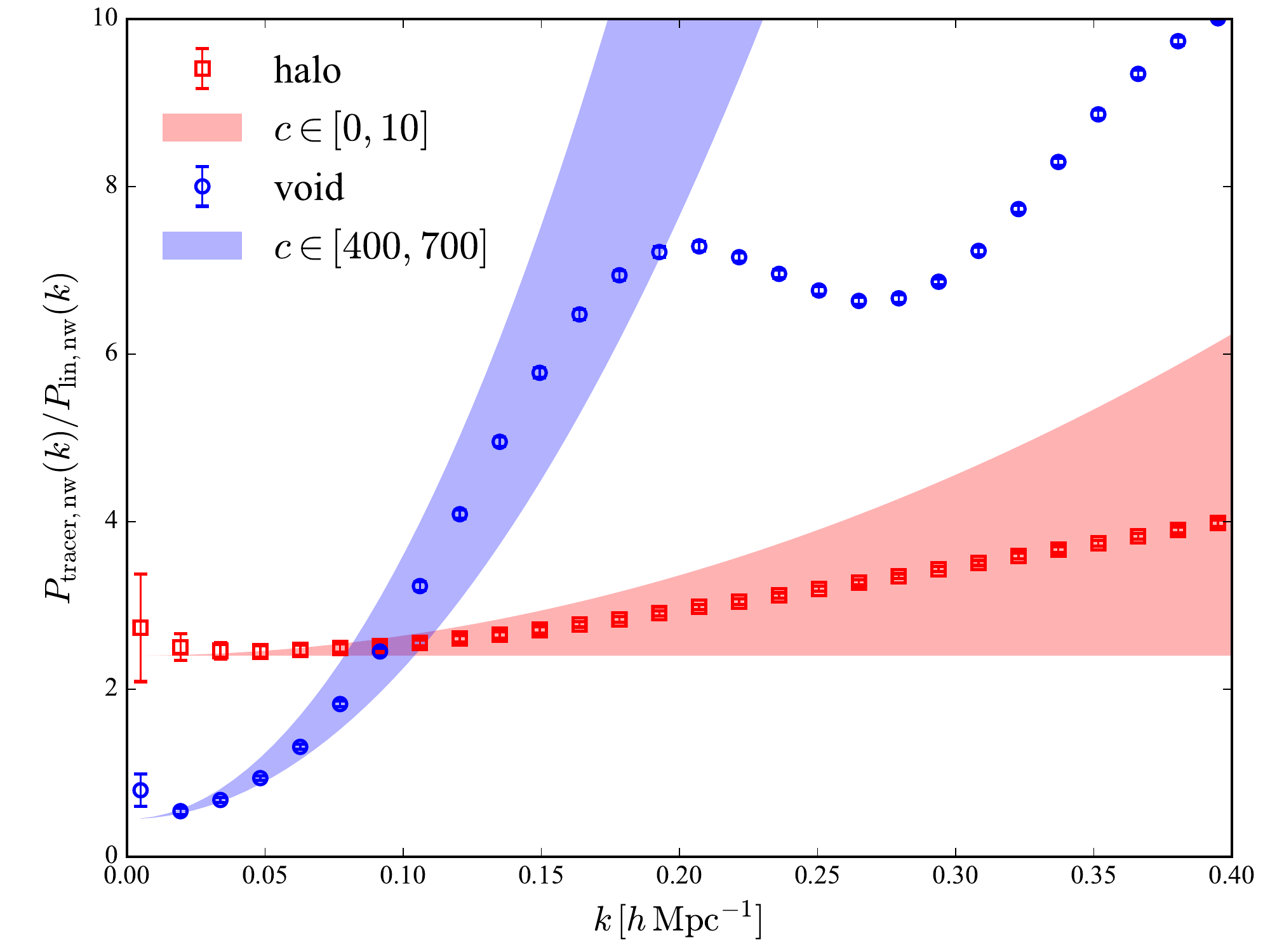}
    \caption{The ratio of the non-wiggle power spectra of haloes and voids measured from \textsc{Patchy} mocks to that of linear dark matter (dashed lines), as well as parabolas $c_0 (1 + c k^2)$ with different values of $c$ (shadowed areas). The error bars show the standard deviation obtained from 100 mocks. For the red area, $c_0 = 2.4$, while for the blue area $c_0 = 0.45$. Therefore, the curve for haloes can be approximated by a constant, but for voids a constant is not enough for the modelling.
}
    \label{fig:nw_ratio}
\end{figure}

To parametrize this extra term in the new BAO model, we plot the measured $P_{\rm tracer, nw} (k) / P_{\rm lin, nw} (k)$ for both haloes and voids from the \textsc{Patchy} cubic mocks in Figure~\ref{fig:nw_ratio}. It shows that the term for haloes is nearly a constant, which explains why the model in Eq.~\ref{eq:pm} works well for haloes. But for voids, the curve is far away from flat. We then use a simple parabola to model this term for wave numbers up to $k \sim 0.2\,h\,{\rm Mpc}^{-1}$:
\begin{equation}
\frac{ P_{\rm t, nw} (k) }{ P_{\rm lin, nw} (k) } = c_0 ( 1 + c k^2 ).
\label{eq:nwratio}
\end{equation}
Since features on larger $k$ are suppressed by the damping factor, they are not important for the BAO pattern (cf. Figure~\ref{fig:bao_ratio}).
Furthermore, considering that the constant factor $c_0$ can be absorbed by the parameter $B$ in Eq.~\ref{eq:xim}, we can rewrite Eq.~\ref{eq:pm1} as
\begin{equation}
\begin{aligned}
&P_{\rm t} (k) = \\
&\quad \left[ ( P_{\rm lin} (k) - P_{\rm lin, nw} (k) ) \, {\rm e}^{-k^2 \Sigma_{\rm nl}^{\prime 2} / 2} + P_{\rm lin, nw} (k) \right] \cdot (1 + c k^2) .
\end{aligned}
\label{eq:pkm}
\end{equation}

%Moreover, since the linear non-wiggle matter power spectrum $P_{\rm lin, nw} (k)$ is a smooth function, the factor $(1 + c k^2)$ is coupled with the nuisance polynomial parameters. We can further simplify the BAO model as
%\begin{equation}
%P_{\rm t} (k) = ( P_{\rm lin} (k) - P_{\rm lin, nw} (k) ) (1 + c k^2) \, {\rm e}^{-k^2 \Sigma_{\rm nl}^{\prime 2} / 2} + P_{\rm lin, nw} (k)
%\label{eq:pkm}
%\end{equation}

With this additional $c$ parameter, we have now 7 free parameters for our BAO model, i.e., $\alpha, B, \Sigma_{\rm nl}, a_1, a_2, a_3$, and $c$.

\subsection{Fit results with the modified de-wiggled model}
The shadowed areas in Figure~\ref{fig:nw_ratio} indicate that the $c$ parameter for haloes is close to 0, while it is a few hundred for voids. Thus, we choose a flat prior of $c$ from $c_1 = -100\,h^{-2}\,{\rm Mpc}^2$ to $c_2 = 900\,h^{-2}\,{\rm Mpc}^2$, and Gaussian tails with a width of $\sigma_c = 100\,h^{-2}\,{\rm Mpc}^2$ for $c$ less than $c_1$ or larger than $c_2$:
\begin{equation}
p(c | {\rm model}) =
\begin{cases}
0, & c < -400\,h^{-2}\,{\rm Mpc}^2 \\
\exp{\left( - \frac{(c - c_1)^2}{2 \sigma_c^2} \right)}, & c \in [-400, -100)\,h^{-2}\,{\rm Mpc}^2 \\
1, & c \in [-100, 900]\,h^{-2}\,{\rm Mpc}^2 \\
\exp{\left( - \frac{(c_2 - c)^2}{2 \sigma_c^2} \right)}, & c \in (900, 1200)\,h^{-2}\,{\rm Mpc}^2 \\
0, & c > 1200\,h^{-2}\,{\rm Mpc}^2
\end{cases}
\end{equation}

\begin{figure}
    \centering
    \includegraphics[width=\columnwidth]{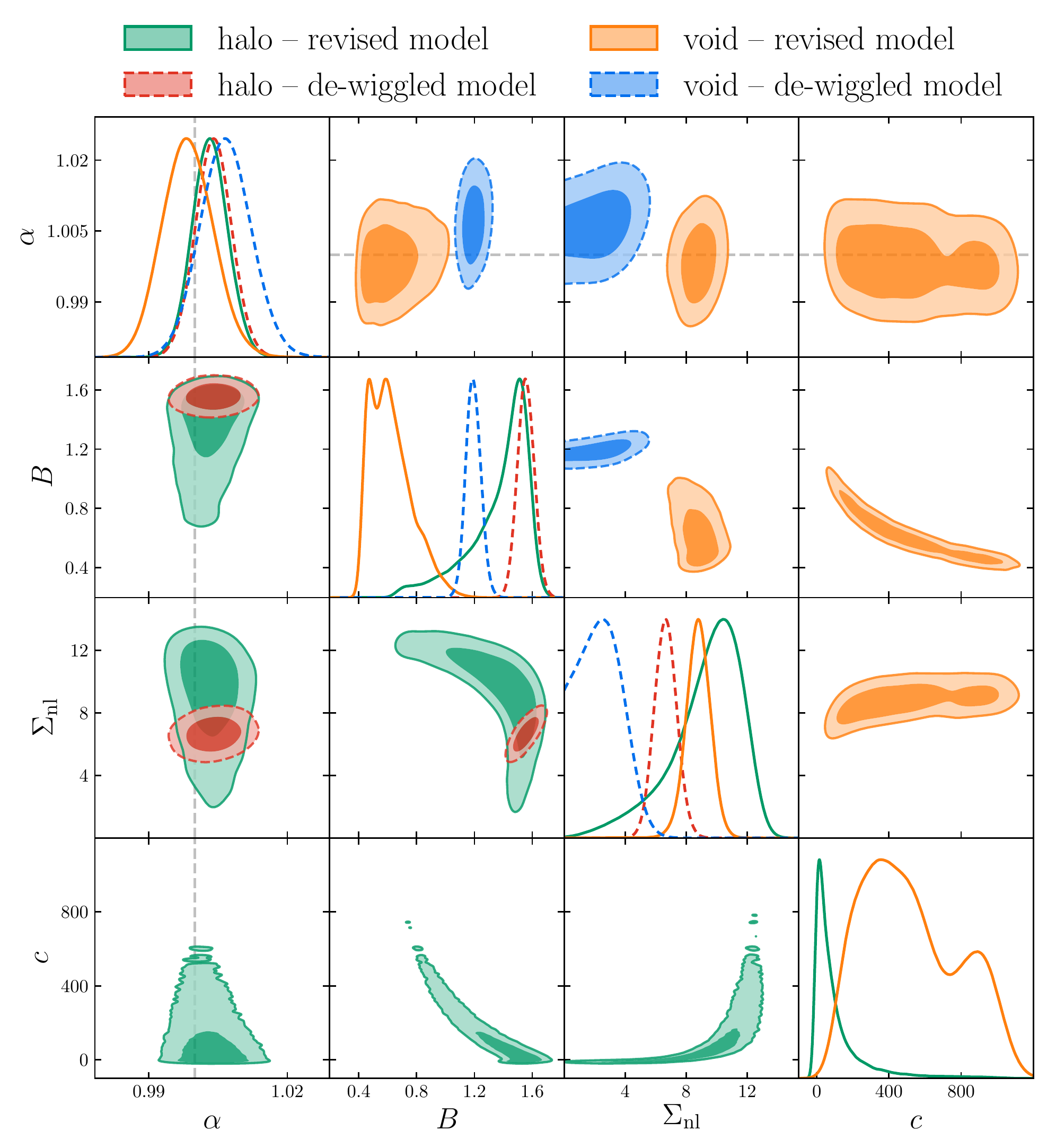}
    \caption{Comparison of the posterior distribution of parameters between the de-wiggled model and the revised model indicated by Eq.~\ref{eq:pkm}, from the fits to the mean 2PCF of 100 pre-reconstruction \patchy{} cubic mock catalogues. The dashed lines indicate the expected $\alpha$ given the fiducial cosmology.}
    \label{fig:triangle_box_new_model}
\end{figure}

\begin{figure*}
    \centering
    \includegraphics[width=.9\columnwidth]{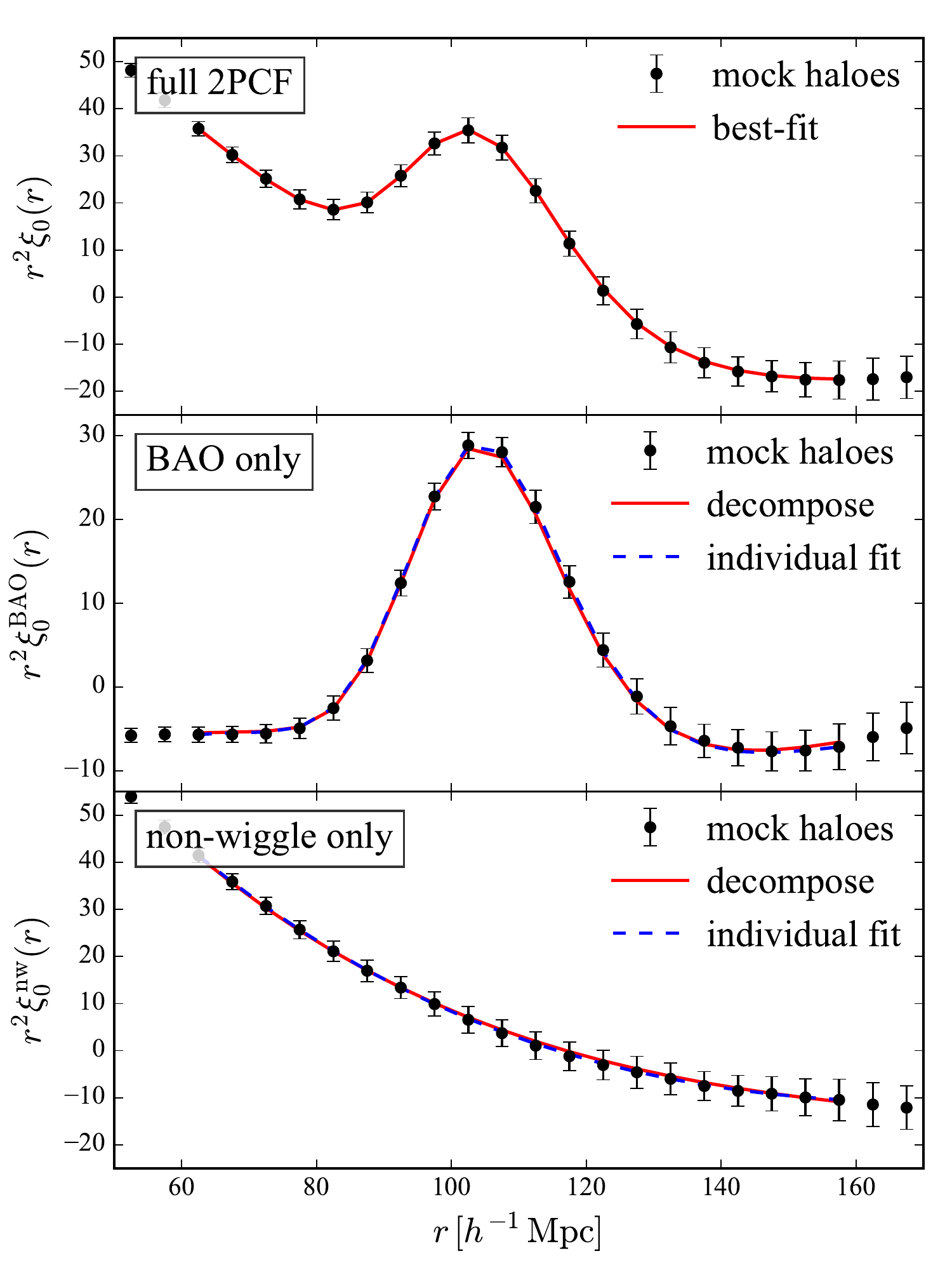}
    \includegraphics[width=.9\columnwidth]{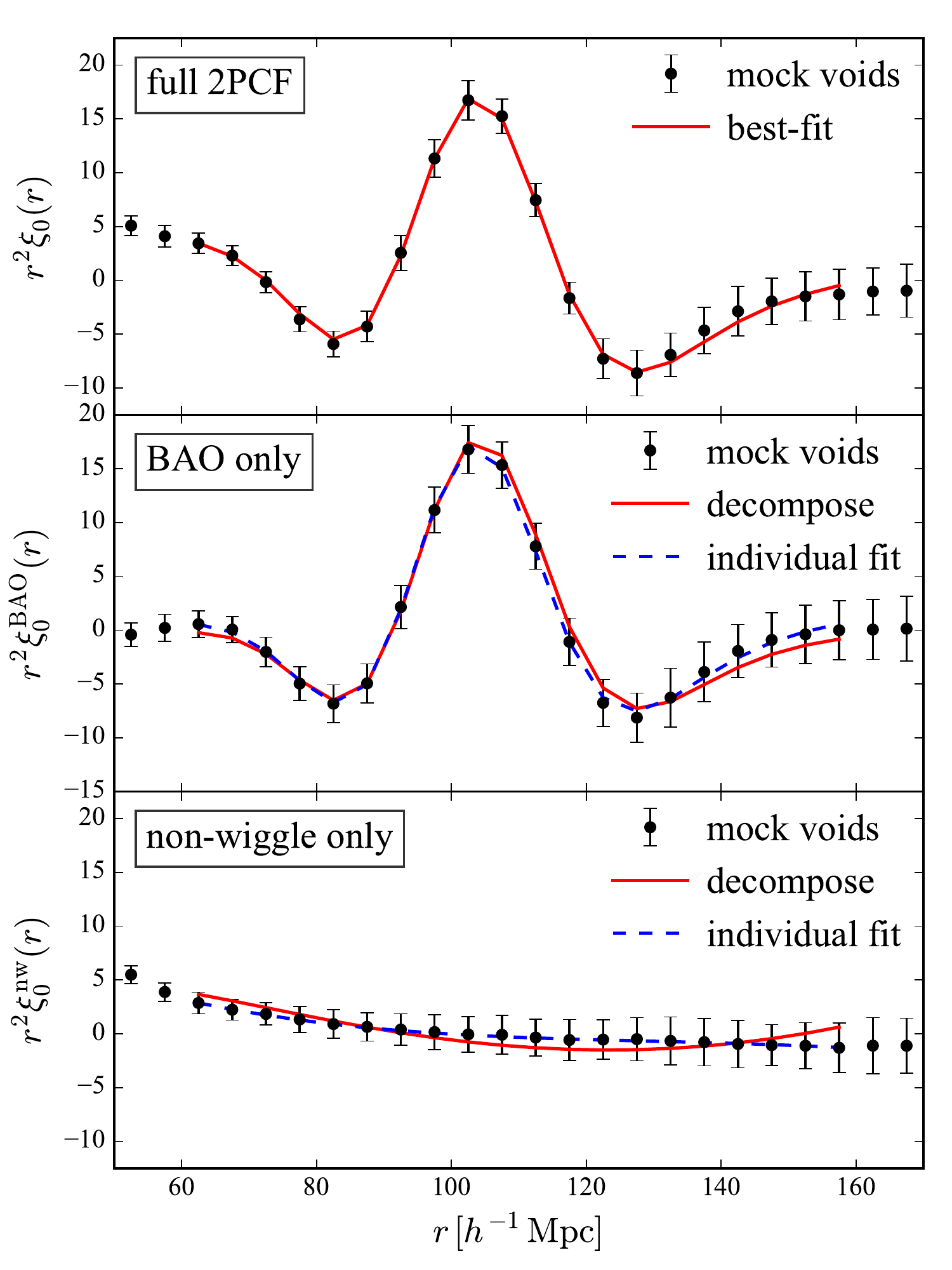}
    \caption{The mean and error of the 2PCFs from 100 Patchy mock {\it halo} ({\it left}) and {\it void} ({\it right}) catalogues (black dots) as well as the best-fit model curves, and the decomposed best-fit curves obtained from the improved BAO model (Eq.~\ref{eq:pkm}). See Figure~\ref{fig:bestfit_box_orig_model} for a detailed description of the panels.}
    \label{fig:bestfit_box_new_model}
\end{figure*}

With this prior we redo the analyses in \S\ref{sec:res1} replacing the BAO model with the one expressed in Eq.~\ref{eq:pkm}.
The posterior of parameters and best-fit (maximum likelihood) model curves for haloes and voids are shown in Figure~\ref{fig:triangle_box_new_model} and Figure~\ref{fig:bestfit_box_new_model} respectively. Though the posterior of the $c$ parameter is obviously non-Gaussian, the maximum likelihood model curves for voids are much closer to the data compared to Figure~\ref{fig:bestfit_box_orig_model}, and the best-fit model for haloes is still consistent with the data. Furthermore, despite the fact that the posterior distribution of both $B$ and $\Sigma_{\rm nl}$ parameters for the haloes are distorted, the fitted $\alpha$ parameter is almost unchanged. It indicates that the $c$ parameter is tightly coupled with $B$ and $\Sigma_{\rm nl}$, but does not affect $\alpha$ significantly. Moreover, the extra parameter improves the fit for voids, and yields better $\alpha$ constraint. In this case, it is still safe to use this revised model to evaluate the BAO peak position.

\begin{table}
\caption{
Fit parameters for the mean 2PCF of 100 \textsc{Patchy} cubic mocks with the improved BAO model expressed by Eq.~\ref{eq:pkm}. The confidence levels are drawn from the posterior shown in Figure~\ref{fig:triangle_box_new_model} and Figure~\ref{fig:triangle_box_post}. And $\mathcal{Z}$ denotes the Bayesian evidence.
}
\centering
\begin{tabular}{ccc}
\toprule
Parameter & halo & void \\
\midrule
 & \multicolumn{2}{c}{pre-reconstruction} \\
$\alpha - 1$ & $0.00329 \pm 0.00381$ & $-0.00152 \pm 0.00553$ \\
$B$ & $1.45 \pm 0.158$ & $0.57 \pm 0.132$ \\
$\Sigma_{\rm nl}$ & $10.07 \pm 2.048$ & $8.80 \pm 0.774$ \\
$c$ & $46.6 \pm 64.9$ & $528.7 \pm 381.0$ \\
$\ln{\mathcal{Z}}$ & $-14.6 \pm 0.15$ & $-14.0 \pm 0.09$ \\
$\chi_{\rm min}^2$ & 0.184 & 1.72 \\
\midrule
 & \multicolumn{2}{c}{post-reconstruction} \\
$\alpha - 1$ & $0.00070 \pm 0.00277$ & $-0.00162 \pm 0.00445$ \\
$B$ & $1.54 \pm 0.067$ & $0.75 \pm 0.174$ \\
$\Sigma_{\rm nl}$ & $6.35 \pm 2.403$ & $9.43 \pm 0.722$ \\
$c$ & $7.6 \pm 16.6$ & $455.0 \pm 285.1$ \\
$\ln{\mathcal{Z}}$ & $-16.5 \pm 0.05$ & $-13.4 \pm 0.08$ \\
$\chi_{\rm min}^2$ & 0.136 & 0.984 \\
\bottomrule
% \toprule
% \multirow{2}{*}{Parameter} & \multicolumn{2}{c}{pre-recon} & \multicolumn{2}{c}{post-recon} \\
% & Halo & Void & Halo & Void \\
% \midrule
% $\bar{\alpha} - 1$ & 0.00394 & $-0.00159$ & 0.00060 & $-0.00110$ \\
% $\sigma_\alpha$ & 0.00389 & 0.00538 & 0.00276 & 0.00469 \\
% $\Sigma_{\rm nl}^\prime$ & 6.62 & 9.28 & 4.54 & 9.67 \\
% $c$ & 0.180 & 20126 & 0.00243 & 973.9 \\
% $B$ & 1.55 & 0.104 & 1.56 & 0.589 \\
% $\chi_{\rm min}^2$ & 0.198 & 1.09 & 0.173 & 0.703 \\
% \bottomrule
\end{tabular}
\label{tab:fit_new_model}
\end{table}

We further list the fitted parameters in Table~\ref{tab:fit_new_model}. For haloes, the fitted parameter $c$ is consistent with 0. The best-fit curve in the left panel of Figure~\ref{fig:bestfit_box_new_model} is very similar to the corresponding one in Figure~\ref{fig:bestfit_box_orig_model}, i.e., the new model degenerates to the de-wiggled model. It also implies that the extra parameter does not improve the modelling of the halo 2PCF, thus the Bayesian evidence $\mathcal{Z}$ is greatly reduced.

In contrast, the best-fit $c$ parameter is fairly large for voids, and in this case the Bayesian evidence is much larger than the value in Table~\ref{tab:fit_orig_model}. Moreover, the fitted $\alpha$ parameter for voids changes more significantly.
% indicating that the $\alpha$ value obtained using the de-wiggled model is biased by the poor fitting.
Therefore, the extra parameter $c$ is crucial for the BAO scale measurement of voids, though it raises degeneracy between parameters.

\begin{figure}
    \centering
    \includegraphics[width=\columnwidth]{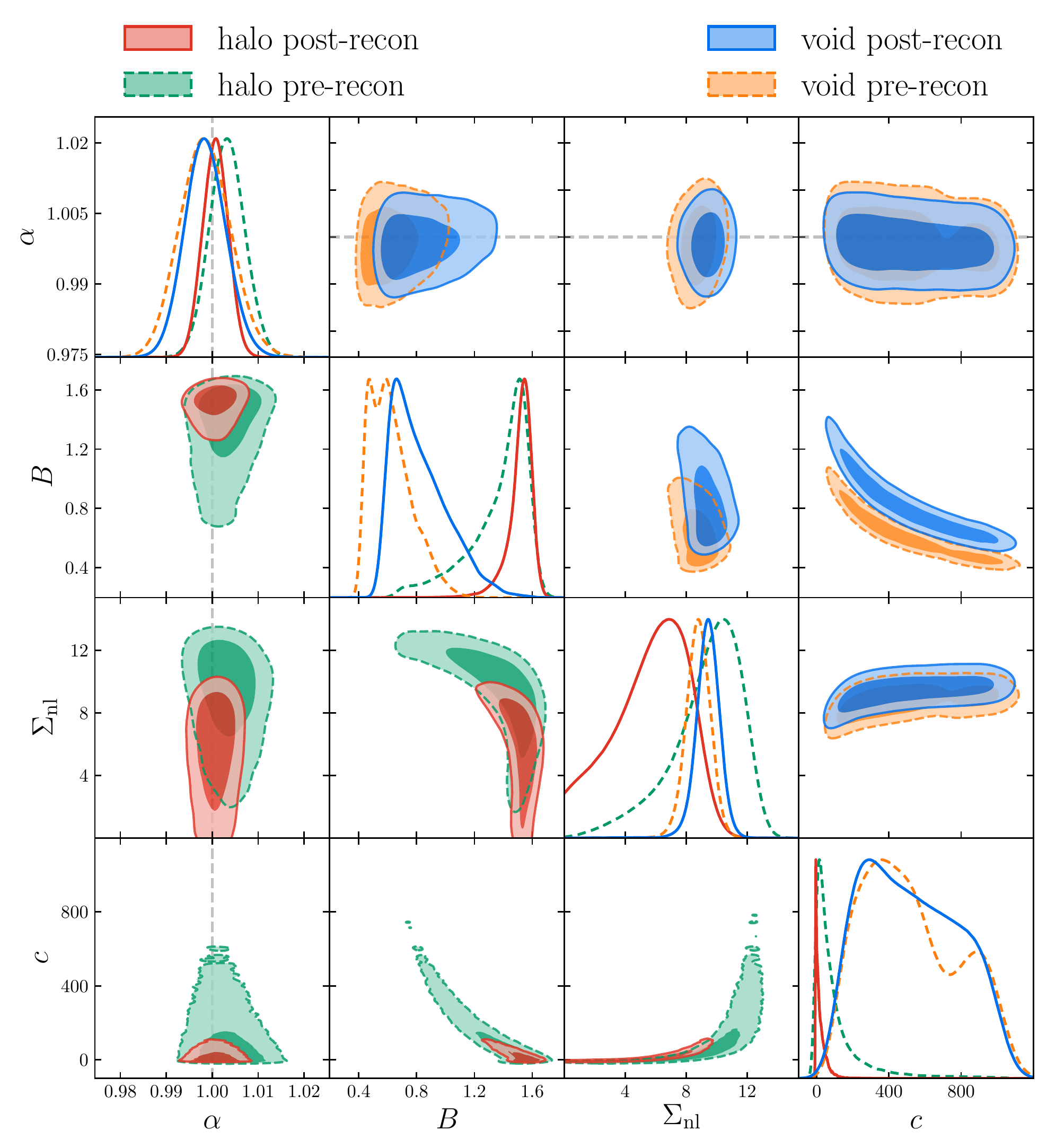}
    \caption{Comparison of the posterior distribution of parameters between the post-reconstruction and pre-reconstruction \patchy{} mocks, from the fits to the mean 2PCF of 100 realisations. The dashed lines indicate the expected $\alpha$ value.}
    \label{fig:triangle_box_post}
\end{figure}

\begin{figure}
    \centering
    \includegraphics[width=\columnwidth]{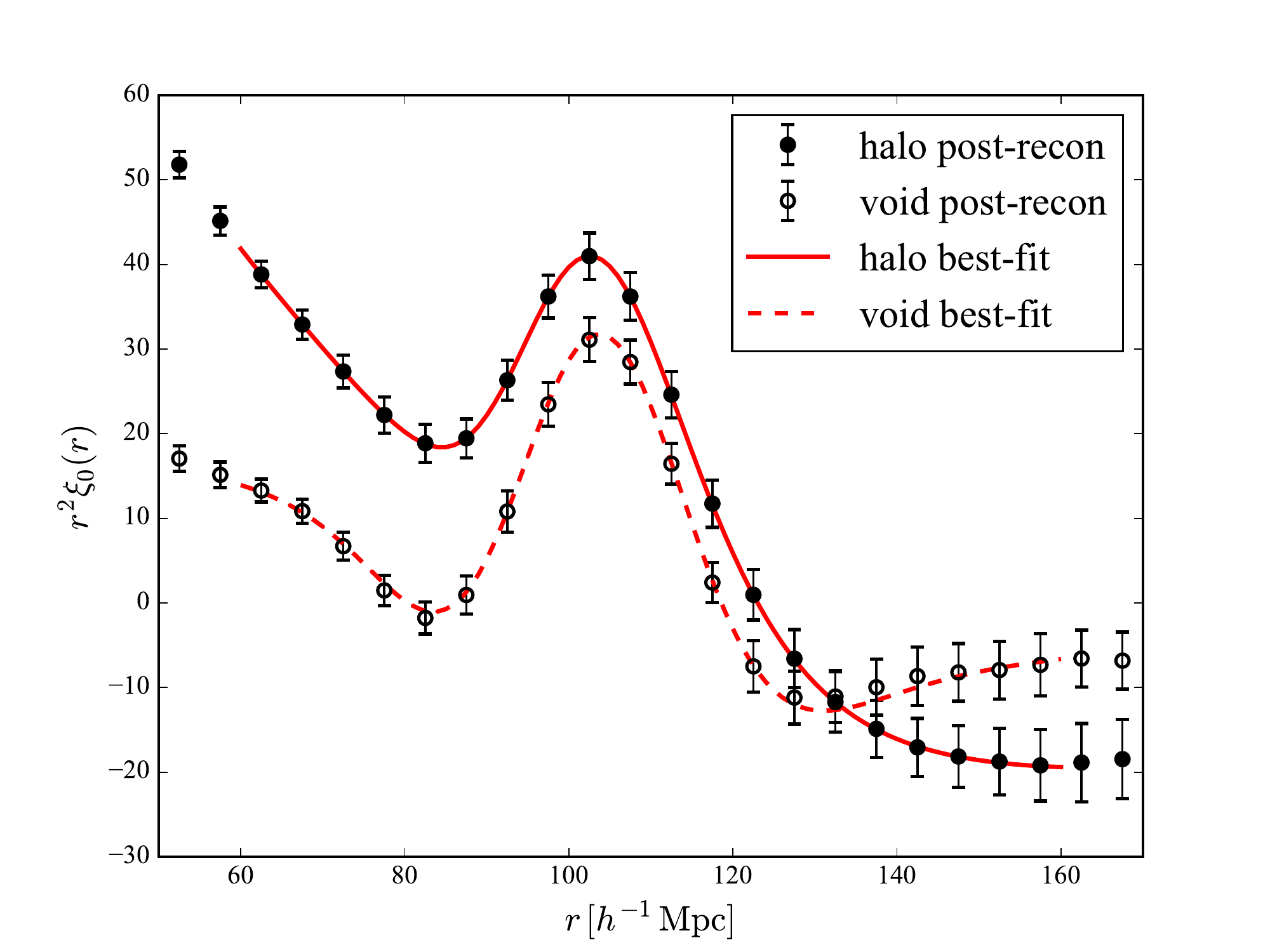}
    \caption{The mean and error of the 2PCFs from 100 \textsc{Patchy} post-reconstruction mock halo/void catalogues (black solid/open dots) as well as the best-fit (maximum likelihood) model curves (red solid/dashed lines) obtained from the modified BAO model.}
    \label{fig:post_new_model}
\end{figure}

We now examine the BAO fit results for post-reconstruction mocks with the modified BAO model.
The posterior distributions and best-fit curves are shown in Figure~\ref{fig:triangle_box_post} and Figure~\ref{fig:post_new_model}, respectively. And the fit results are listed in Table~\ref{tab:fit_new_model}.
In this case, the Bayesian evidence for haloes is further reduced, this is due to the fact that a non-zero $c$ parameter is strongly disfavoured by the post-reconstruction halo 2PCF. This is not a problem, since the fitted $\alpha$ is not obviously biased by the introduction of the $c$ parameter. Indeed, the posterior distribution of $\alpha$ is almost orthogonal to that of all the other parameters. Besides, the maximum likelihood model curves in Figure~\ref{fig:post_new_model} are in good agreement with the 2PCF measured from mocks.
 
A further improvement of the BAO fitting method requires a more detailed modelling of $P_{\rm void, nw} (k) / P_{\rm lin, nw} (k)$ (see Eq.~\ref{eq:nwratio}). However, this would complicate the model and introduce more fitting parameters. Our main goal is to explore a precise measurement of the BAO scale with a combination of galaxies and voids, rather than constructing a perfect BAO model for voids only. Therefore, we consider the void BAO fit good enough within the scope of this paper.

\subsection{Discussions on the BAO fit results}

The fit results in Table~\ref{tab:fit_new_model} show that BAO reconstruction does yield smaller errors on $\alpha$ for both haloes and voids.
Basically we obtain a $\sim 25\,\%$ reduction of the error on the baryon acoustic scale constraint for haloes after BAO reconstruction, while the improvement on the constraint is $\sim 20\,\%$ for voids. Besides, the BAO scale constraining power of voids is weaker than that of haloes, for both pre- and post-reconstruction samples. This is because voids are indirect tracers based on higher order statistics of the halo distribution, which encodes less information on the linear density field compared to the 2PCF of haloes.
Moreover, after reconstruction, the 2PCF of haloes get closer to the linear one, so the additional (improved) information from voids should be further reduced, compared to the case without reconstruction.

Apart from $\sigma_\alpha$, the fitted mean of the $\alpha$ parameter also varies for different samples. Since the fiducial cosmology we use for the fits in this section is the `true' cosmology for generating the mocks, the fitted $\alpha$ is expected to be exactly 1.
For the pre-reconstruction case, Table~\ref{tab:fit_new_model} shows opposite BAO shifts for haloes and voids, which is consistent with the density-dependent BAO peak motions shown in theoretical and numerical studies \citep[e.g.][]{McCullagh2013, Neyrinck2018}.
Note that the covariance matrix we use here indicates the error of a single realisation, but the fitted data is the mean 2PCF of 100 mocks. Therefore, considering only sample variance, the fitted error of the mean 2PCF should be $\sim 10\,\%$ of that of a single realisation. In this case, the BAO shifts are statistically significant ($\sim 10\,\sigma$ for haloes and $\sim 3\,\sigma$ for voids).

In brief, over-densities collapse and pull the BAO peak inward, while under-densities expand and push the peak outward.
Since the BAO shift is mainly caused by the non-linear gravitational evolution of over-densities, the weak gravity of under-dense regions explains why the shift of voids is much smaller than that of haloes. As a result, the evolution of voids are more consistent with linear theories, thus the 2PCF of voids are closer to the linear one on BAO scales.
This further confirms that the BAO signal of our void sample is from under-densities, and the revised model is able to extract this feature.

The BAO shift of haloes is reduced by 80\,\% after BAO reconstruction, this makes the measurement of $\alpha$ from haloes closer to the expected value. However, the BAO shift of voids is slight larger, this may be due to the small movement of under-densities over cosmic time \citep[][]{Sutter2014b}, thus, the displacement of troughs are relatively low, compared to statistical noises. Nevertheless, after reconstruction, the BAO peak position measured from haloes are in better agreement, thus a joint baryon acoustic scale constraint with haloes (or galaxies) and voids is more likely to be unbiased.

We shall focus on the post reconstruction case for the rest of this paper, since our main goal is to explore the possibility of improving the BAO measurement using voids, on top of the BAO reconstruction method.
However, it is possible that the optimal radius cut for the post-reconstruction catalogues is different from the one measured by \citet[][]{Liang2016} using pre-reconstruction catalogues, since cosmic voids are smaller at higher redshifts \cite[][]{Wojtak2016}. We shall investigate this effect in future studies.

%%%%%%%%%%%%%%%%%%%%%%

\section{Measuring BAO by combining galaxies and voids}
\label{sec:res}

\subsection{Correlation function estimator for the combined sample}
Now we investigate the possibility of constructing a combined galaxy and void sample for baryon acoustic scale constraint.
It has been shown that a population of the voids has negative large-scale bias, and for voids with $R_{\rm V} > 15\,\hmpc$, the BAO peak is negative in the cross correlation function of galaxies and voids \citep[][]{Chuang2017}, while for smaller voids the biases are positive and they are correlated with galaxies, as very small DT voids are actually in over-dense regions.
Since we use mainly the large voids for the BAO measurement in this paper,
if we simply compute the auto-correlation function of the joint data set of the galaxy and void tracers, the BAO signal will be smeared out by adding positive and negative terms together (see the $w=1$ case in Eq.~\ref{eq:cfcomb}).
However, we can use a weighting scheme to assign different weights/signs to galaxies and voids.
%Therefore, putting galaxies and voids together directly is likely to smear out the BAO signal. Instead, we can use a weighting scheme, and assign some weights to galaxies and voids.

In this paper, we consider the simplest weighting scheme, i.e. assigning a constant weight $w$ to the selected void tracers, while leaving galaxies without extra weights.
In this case, a galaxy-void pair and a void-void pair will be counted as $w$ and $w^2$ pairs, respectively.
Denoting the normalised counting pairs for galaxies (voids) with subscript g (v), and the number of galaxies and voids as $n_{\rm g}$ and $n_{\rm v}$ respectively. Since the un-normalised pair counts and the number of tracers should be added, the combined $DD$ term with all the voids being weighted by $w$ is
\begin{equation}
\begin{aligned}
DD &= \frac{D_{\rm g} D_{\rm g} \cdot ( n_{\rm g}^2 / 2) + D_{\rm g} D_{\rm v} \cdot n_{\rm g} n_{\rm v} w + D_{\rm v} D_{\rm v} \cdot (n_{\rm v}^2 / 2) w^2}{(n_{\rm g} + n_{\rm v} w)^2 / 2} \\
&= \frac{D_{\rm g} D_{\rm g} \cdot n_{\rm g}^2 + 2 D_{\rm g} D_{\rm v} \cdot n_{\rm g} n_{\rm v} w + D_{\rm v} D_{\rm v} \cdot n_{\rm v}^2 w^2}{(n_{\rm g} + n_{\rm v} w)^2} .
\end{aligned}
\end{equation}
The factors $1/2$ in the first equation eliminate self-counting.
Similarly, the $DR$ and $RR$ terms for the combined sample are
\begin{align}
DR &= \frac{ D_{\rm g} R_{\rm g} \cdot n_{\rm g}^2 + (D_{\rm g} R_{\rm v} + D_{\rm v} R_{\rm g}) \cdot n_{\rm g} n_{\rm v} w + D_{\rm v} R_{\rm v} \cdot n_{\rm v}^2 w^2 }{ (n_{\rm g} + n_{\rm v} w)^2 } ,\\
RR
&= \frac{R_{\rm g} R_{\rm g} \cdot n_{\rm g}^2 + 2 R_{\rm g} R_{\rm v} \cdot n_{\rm g} n_{\rm v} w + R_{\rm v} R_{\rm v} \cdot n_{\rm v}^2 w^2}{(n_{\rm g} + n_{\rm v} w)^2} .
\end{align}
Note that the normalisation factors in these equations are expressed by $n_{\rm g}$ and $n_{\rm v}$, instead of the corresponding number of random points. This is because throughout this paper we use the random catalogues containing 20 times more objects than the data catalogues, for both galaxies and voids, thus the factor 20 is eliminated.

The correlation function of the combined sample is then
\begin{equation}
\begin{aligned}
\xi_{\rm comb}
&= \frac{DD - 2DR + RR}{RR} \\
&= \frac{n_{\rm g}^2 \xi_{\rm gg} \cdot R_{\rm g} R_{\rm g} + 2 n_{\rm g} n_{\rm v} w \xi_{\rm gv} \cdot R_{\rm g} R_{\rm v}  + n_{\rm v}^2 w^2 \xi_{\rm vv} \cdot R_{\rm v} R_{\rm v}}{n_{\rm g}^2 \cdot R_{\rm g} R_{\rm g} + 2 n_{\rm g} n_{\rm v} w \cdot R_{\rm g} R_{\rm v} + n_{\rm v}^2 w^2 \cdot R_{\rm v} R_{\rm v}} ,
\end{aligned}
\label{eq:cfcomb}
\end{equation}
where
\begin{subequations}
\begin{align}
\xi_{\rm gg} &= \frac{D_{\rm g} D_{\rm g} - 2 D_{\rm g} R_{\rm g} + R_{\rm g} R_{\rm g}}{R_{\rm g} R_{\rm g}} ,\\
\xi_{\rm vv} &= \frac{D_{\rm v} D_{\rm v} - 2 D_{\rm v} R_{\rm v} + R_{\rm v} R_{\rm v}}{R_{\rm v} R_{\rm v}} ,\\
\xi_{\rm gv} &= \frac{D_{\rm g} D_{\rm v} - D_{\rm g} R_{\rm v} - D_{\rm v} R_{\rm g} + R_{\rm g} R_{\rm v}}{R_{\rm g} R_{\rm v}} .
\end{align}
\end{subequations}
For the post-reconstruction case, we simply replace the $R_{\rm g}$ term in Eq.~\ref{eq:cfcomb} by $S_{\rm g}$, representing the shifted random catalogue. In fact, it makes essentially no difference to the correlation function, whether using $SS$ or $RR$ (as Eq.~\ref{eq:cfpost}) in the denominator \citep[][]{Padmanabhan2012}.

Since the BAO peak position constraint from voids alone is weaker than that of galaxies alone, the absolute value of the optimal weight for voids should be less than 1. We then vary $w$ in the range $w \in [-1, 1]$ and explore the value that yields the highest precision on the $\alpha$ parameter determination.

\subsection{Optimal combination of galaxies and voids}
\begin{figure}
    \centering
    \includegraphics[width=\columnwidth]{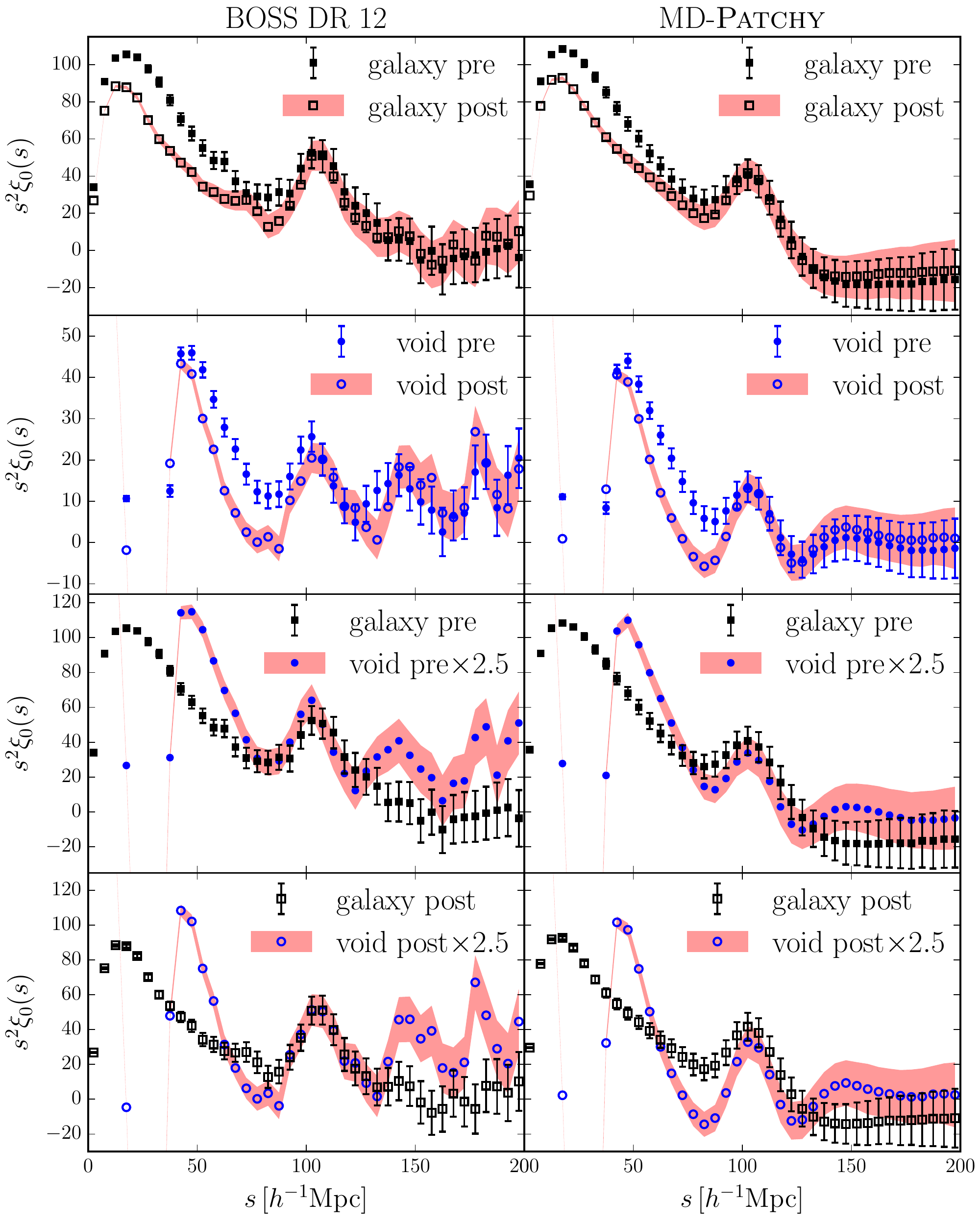}
    \caption{Comparisons of the 2-point correlation functions of pre- and post-reconstruction galaxies and voids in the redshift bin $0.5 < z < 0.75$. `pre' and `post' in the legends are referred to the results from pre-reconstruction and post-reconstruction samples respectively. The error bars and shadowed bands show the error obtained from 1000 realisations of mocks. For the comparisons between galaxies and voids, we multiply the 2PCF of voids by a factor of 2.5 to obtain similar amplitudes on BAO scale.}
    \label{fig:data_cf}
\end{figure}

We rely on the post-reconstruction MD-\patchy{} DR12 light-cone mocks for the optimal void weight search. The 2PCFs of galaxies and voids in the \highz{} bin are shown in Figure~\ref{fig:data_cf}.
Note that the fluctuations on void 2PCFs are visually enhanced on large scales since the vertical axes are scaled by $s^2$, while the actual amplitudes are much less than that of the BAO peak.

\begin{figure}
    \centering
    \includegraphics[width=\columnwidth]{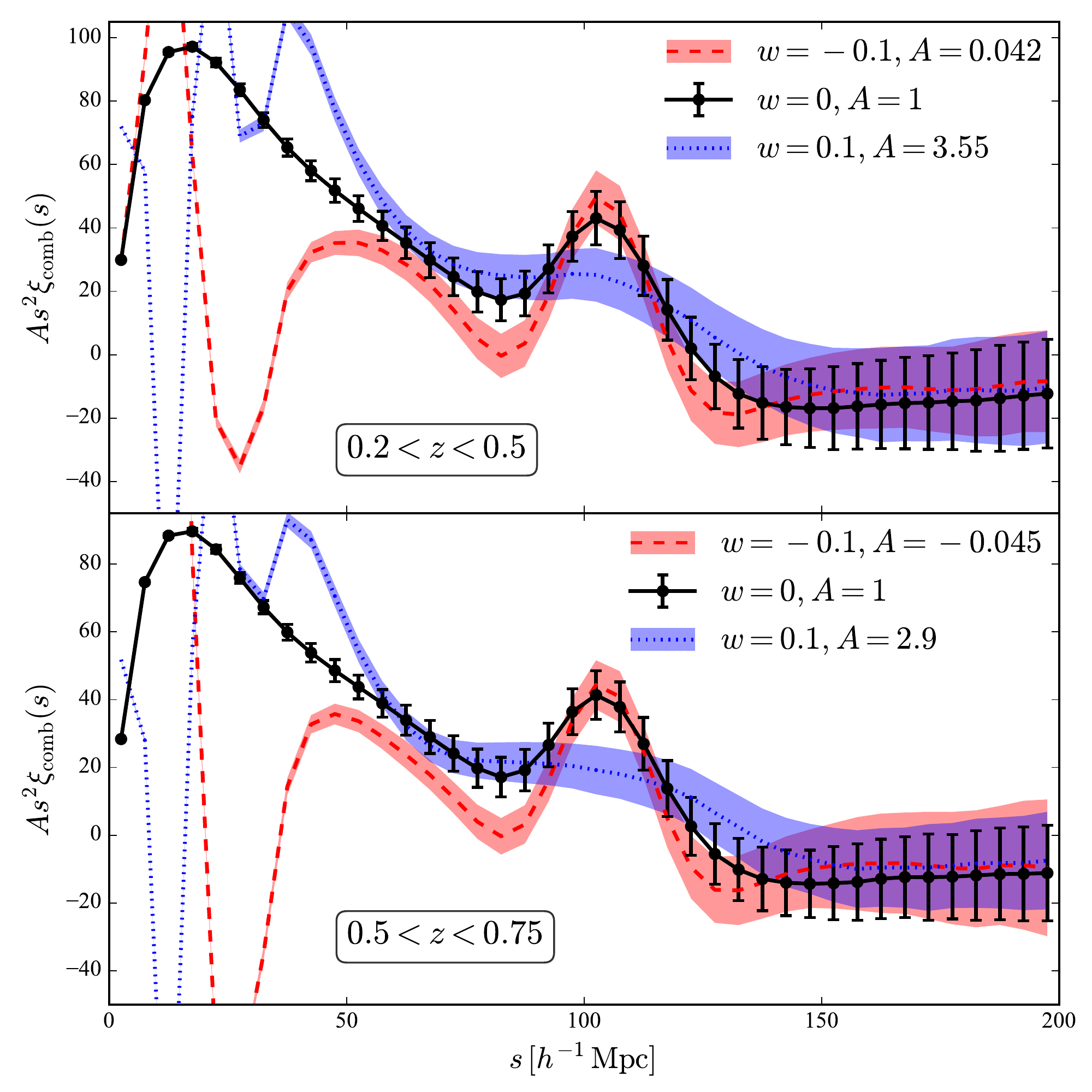}
    \caption{The combined correlation function of post-reconstruction MD-\patchy{} galaxies and voids, with $w$ being $-0.1$, $0$, and $0.1$, for the redshift bins $0.2 < z < 0.5$ ({\it upper}) and $0.5 < z < 0.75$ ({\it lower}), respectively. The error bars and shadowed bands show the error obtained from 1000 realisations of mocks. The curves has been rescaled by constant factors ($A$) to have the same 1\,$\sigma$ error at $s \in [100, 105]\,\hmpc$.}
    \label{fig:cf_weight}
\end{figure}

With the correlation function estimator described in the previous section, the combined correlation function ($\xi_{\rm comb}$) with several $w$ values are shown in Figure~\ref{fig:cf_weight}. In particular, the curves with $w=0$ indicate the galaxy auto-correlation function $\xi_{\rm gg}$.
The amplitude of the combined correlation functions with different $w$ vary significantly, thus, for better visualisation, we have rescaled $\xi_{\rm comb}$ by a constant factor $A$, such that the 1\,$\sigma$ deviation of the curves near the BAO peak ($s \in [100, 105]\,\hmpc$) are the same as that of the galaxy auto-correlation function.
%We have chosen a negative $A$ for the $w=-0.1$ case for the \highz{} sample, as the $\xi_{\rm comb}$ in this case is negative on the BAO scale. Therefore, we flip the curve for a better comparison.

Figure~\ref{fig:cf_weight} shows that for both the \lowz{} and \highz{} bins, the BAO peak with $w=0.1$ is not obvious. This confirms our expectation that a positive $w$ may decrease the significance of the BAO signal. On the contrary, with a weight of $-0.1$, the amplitude of the BAO peak of the combined correlation function is higher than that of $\xi_{\rm gg}$, suggestion a better peak position constraint. To determine the optimal value of $w$, which minimize the standard error of parameter $\alpha$ ($\sigma_\alpha$), we then apply BAO fits to the combined correlation function of MD-\patchy{} mocks with different values of $w$. In this case, we do not rescale $\xi_{\rm comb}$ as in Figure~\ref{fig:cf_weight}. However, since the model expressed by Eq.~\ref{eq:xim} requires a positive BAO peak, we flip the combined correlation function (i.e., fit to $-\xi_{\rm comb}$) when it is negative at around $s \in [90, 110]\,\hmpc$.

Similar to the case of cubic mocks, we fit the BAO model to the mean of the 2PCFs of all the mock realisations. The fitting range we use in this section is $s \in [60, 150]\,\hmpc$.
This range is chosen to have no significant impacts on the fit results, and contains as many data points as possible (see Appendix~\ref{sec:range} for details).

\begin{figure*}
    \centering
    \includegraphics[width=.8\columnwidth]{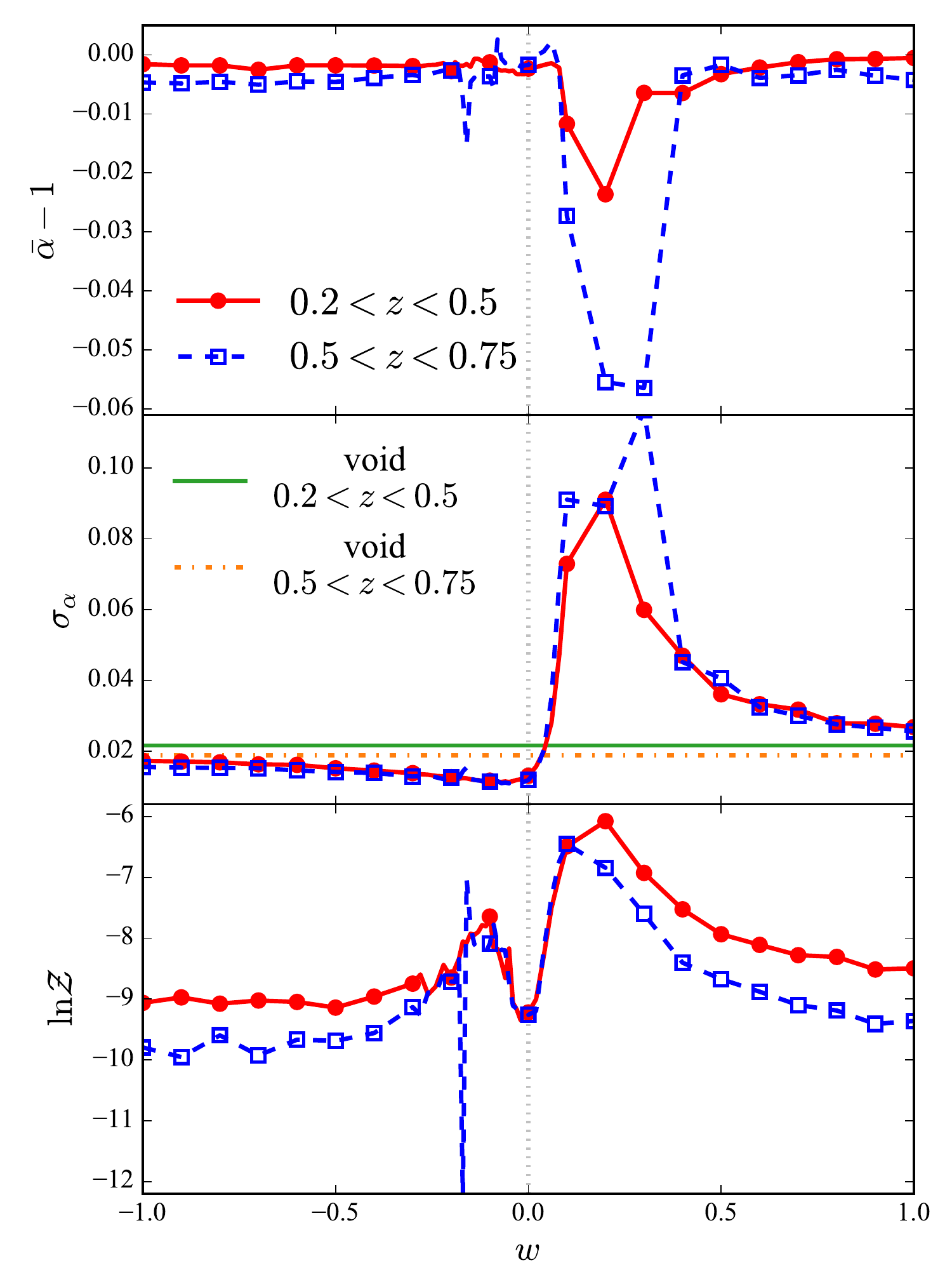}
    \includegraphics[width=.8\columnwidth]{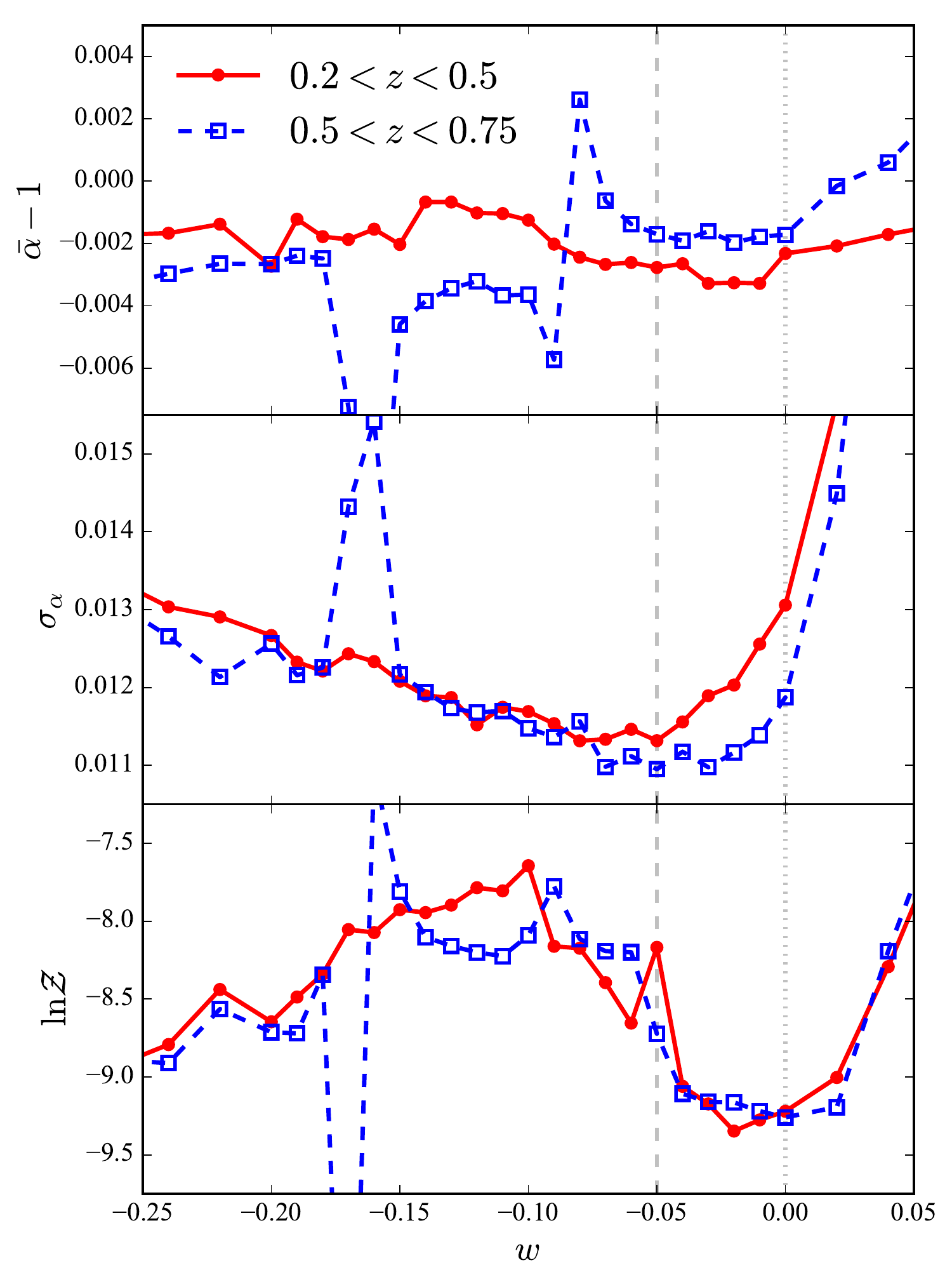}
    \caption{The mean $\alpha - 1$, the error on $\alpha$, as well as the Bayesian evidence as functions of the weight of voids, from fits to the mean of 1000 MD-\patchy{} mocks, in the redshift bins $0.2 < z < 0.5$ (\lowz{}) and $0.5 < z < 0.75$ (\highz{}), respectively. The {\it left} panel shows the results for the full range of $w$ that we have tested, while the {\it right} panel shows only a small $w$ range close to the optimal value ($-0.05$, indicated by the grey dashed line). The grey dotted vertical lines indicate the results from galaxy auto-correlation function ($w=0$), and the horizontal lines in the {\it left} panel shows that fitted $\sigma_\alpha$ from void auto-correlations.}
    \label{fig:best_wt}
\end{figure*}

The fitted mean of $\alpha-1$, $\sigma_\alpha$, and the Bayesian evidence as function of $w$ for the \lowz{} and \highz{} bins are shown in Figure~\ref{fig:best_wt}. We find that for both redshift bins, in the weight range $-0.1 < w < 0$, the combined sample reduces the error on $\alpha$, compared to the results from galaxies alone ($w=0$).
Besides, the $\alpha$ measured from the combined samples are close to the values obtained from galaxies in the same weight range.
Therefore, with a weight in this range, including voids not only improves the BAO peak position measurement, but also keeps it unbiased.
Besides, the Bayesian evidences for $-0.1 < w < 0$ are slightly larger than the value for galaxies, indicating that the model works well in this case.

The optimal weight $w$ for both cases are around $-0.05$, where $\sigma_\alpha$ measured from the combined samples are minimized.
It is also worth noting that when $w$ approaches $\pm 1$, the fitted $\sigma_\alpha$ get closer to that from the void auto-correlation functions ($\xi_{\rm vv}$). It indicates that with $w \sim \pm 1$, $\xi_{\rm vv}$ is already the dominant term in Eq.~\ref{eq:cfcomb}. This can be explained by the fact that $n_{\rm v} \sim 1.5 n_{\rm g}$. Besides, the convergence of $\sigma_\alpha$ also implies that the range of $w$ we have studied is large enough for the exploration of the optimal value.
Furthermore, there are large biases and errors on $\alpha$ when $w \sim 0.2$, this is consistent with the absence of obvious BAO peak for $w=0.1$ shown in Figure~\ref{fig:cf_weight}.

\begin{figure*}
    \centering
    \includegraphics[width=.9\columnwidth]{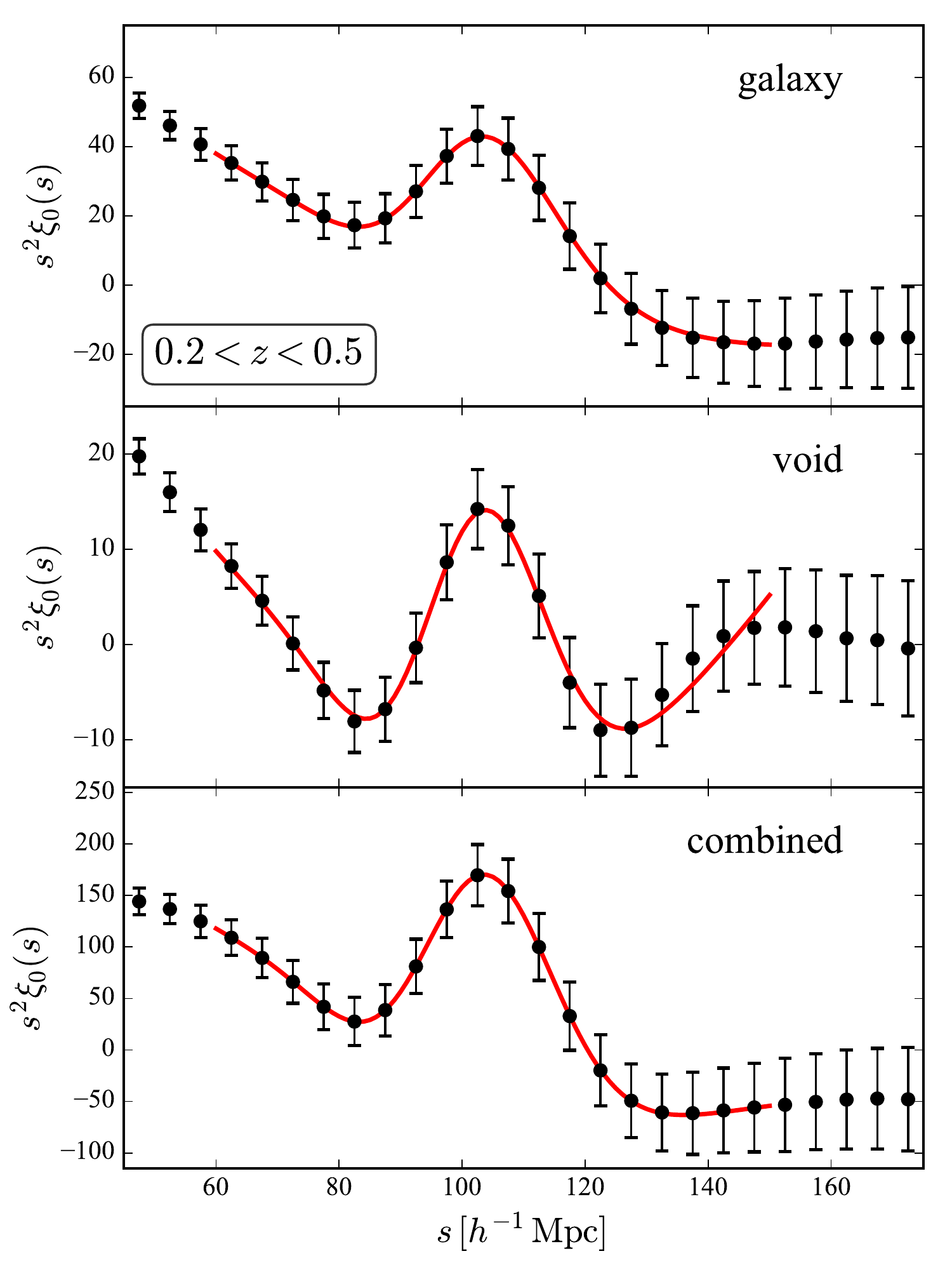}
    \includegraphics[width=.9\columnwidth]{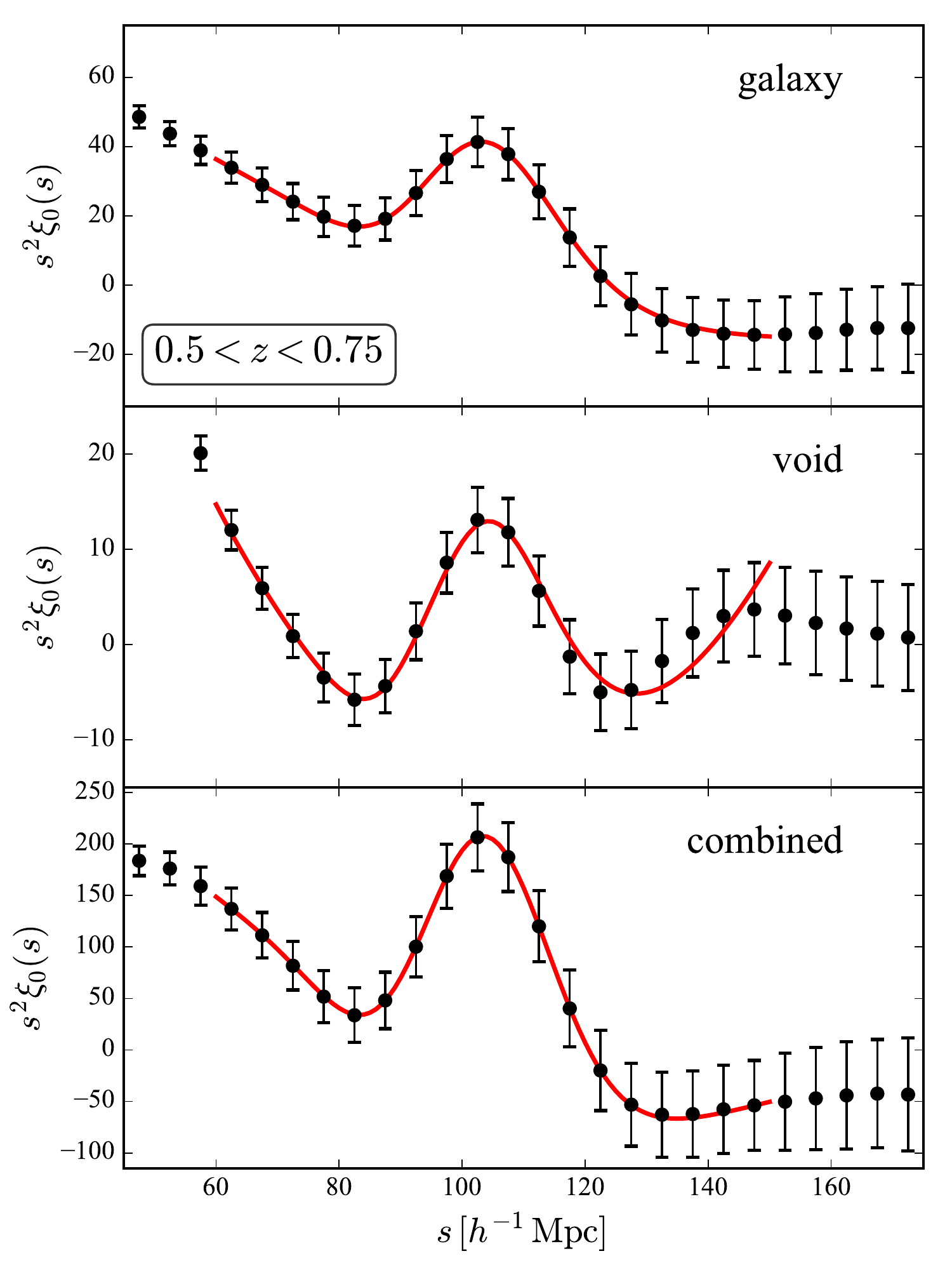}
    \caption{The best-fit curves for the mean of 1000 MD-\patchy{} mocks of the galaxy sample, the void sample, and the combined sample with $w=-0.05$ for the redshift bins $0.2 < z < 0.5$ ({\it left}) and $0.5 < z < 0.75$ ({\it right}), respectively.}
    \label{fig:bestfit_patchy}
\end{figure*}

We present the best-fit (maximum likelihood) curves of the galaxy sample, the void sample, and the combined sample with $w = -0.05$ in Figure~\ref{fig:bestfit_patchy} for the two redshift bins. Indeed, the revised de-wiggled model (Eq.~\ref{eq:pkm}) describes well the 2PCF of the galaxy and combined sample, for the MD-\patchy{} mocks. It implies that the BAO peak scale measured by this model are reliable for these samples. The best-fit curves are not optimal for scales larger than the BAO peak position, but this does not affect our conclusions on the improvement of the BAO scale constraint when voids are included, as we are only interested in the comparison between results from galaxies alone, and the combined sample.

\begin{table}
\caption{The fit results for the mean 2PCF of 1000 MD-\patchy{} light-cone mocks, of the galaxy sample, the void sample, and the combined sample with the void weight $w=-0.05$.}
\centering
\begin{tabular}{ccc}
\toprule
& $0.2 < z < 0.5$ & $0.5 < z < 0.75$ \\
\midrule
& \multicolumn{2}{c}{$\alpha$} \\
galaxy & $0.9977 \pm 0.0131$ & $0.9983 \pm 0.0119$ \\
void & $0.9992 \pm 0.0216$ & $0.9957 \pm 0.0189$ \\
combined & $0.9972 \pm 0.0113$ & $0.9983 \pm 0.0110$\\
\midrule
& \multicolumn{2}{c}{$\ln{\mathcal{Z}}$} \\
galaxy & $-9.2 \pm 0.10$ & $-9.3 \pm 0.13$ \\
void & $-9.0 \pm 0.06$ & $-9.8 \pm 0.05$ \\
combined & $-8.2 \pm 0.36$ & $-8.7 \pm 0.05$ \\
\midrule
& \multicolumn{2}{c}{$\chi_{\rm min}^2$} \\
galaxy & 0.046 & 0.040 \\
void & 0.70 & 1.6 \\
combined & 0.014 & 0.029 \\
% & \multicolumn{2}{c}{$\alpha \pm \sigma_\alpha$} \\
% galaxy & $0.9981 \pm 0.0133$ & $0.9996 \pm 0.0124$ \\
% void & $0.9963 \pm 0.0202$ & $1.0178 \pm 0.0575$ \\
% combined & $0.9982 \pm 0.0114$ & $0.9998 \pm 0.0110$\\
% \midrule
% & \multicolumn{2}{c}{$\chi_{\rm min}^2$} \\
% galaxy & 0.128 & 0.143 \\
% void & 1.79 & 9.47 \\
% combined & 0.127 & 0.132 \\
\bottomrule
\end{tabular}
\label{tab:fit_lightcone}
\end{table}

The fit results for the mean of 1000 MD-\patchy{} mocks are listed in Table~\ref{tab:fit_lightcone}.
The numbers for the galaxy samples are consistent with the results in \citet[][]{Vargas2018}, with the same sets of mocks.
Qualitatively, the negative optimal $w$ maximises the correlation signal expressed in Eq.~\ref{eq:cfcomb}, by both increasing the numerator and reducing the denominator, as the galaxy-void cross correlation is negative on BAO scales \citep[cf.][]{Chuang2017}. Besides, the small absolute value of $w$ `down-weights' the void sample size $n_{\rm v}$. It indicates that the `effective' number of void tracers that contribute to cosmological constraint is actually much smaller than that of galaxies, despite the fact that we have more DT voids than galaxies. This is also consistent with the larger $\sigma_\alpha$ from voids in Table~\ref{tab:fit_lightcone}. In any case, a non-zero $w$ indicates that there is still contribution from voids.
Indeed, using the combined sample, we obtain $13.7\,\%$ and $9.8\,\%$ improvement over the galaxy sample alone, for the precision of baryon acoustic scale constraints in the \lowz{} bin and \highz{} bin respectively, by comparing the fitted $\sigma_\alpha$: $( \sigma_{\alpha, {\rm combined}} - \sigma_{\alpha, {\rm galaxy}} ) / \sigma_{\alpha, {\rm galaxy}}$. These amounts of improvement correspond to effectively enlarging the sample by more than 20\,\%.

Therefore, voids, or troughs of the underlying density field, do encode additional (and complementary) cosmological information compared to galaxies, even with the existing BAO reconstruction technique. Thus, even though BAO reconstruction transfers higher order statistical information from the distribution of galaxies to the two point statistics by linearising the position of the tracers \citep[][]{Schmittfull2015}, there is still information that can be gained by extracting the clustering in under-dense regions. This is because the information from under-densities are lost when measuring only the clustering of galaxies, which is similar to a truncation of the Gaussian field.
Our method restores more information from the truncated under-density regions.
In addition, to examine the robustness of the conclusion in terms of the number of mocks, we further divide our mock sample into two independent sub-samples with 500 mocks, and both of the sub-samples yield the same result as the full sample.

\begin{figure}
\centering
\includegraphics[width=\columnwidth]{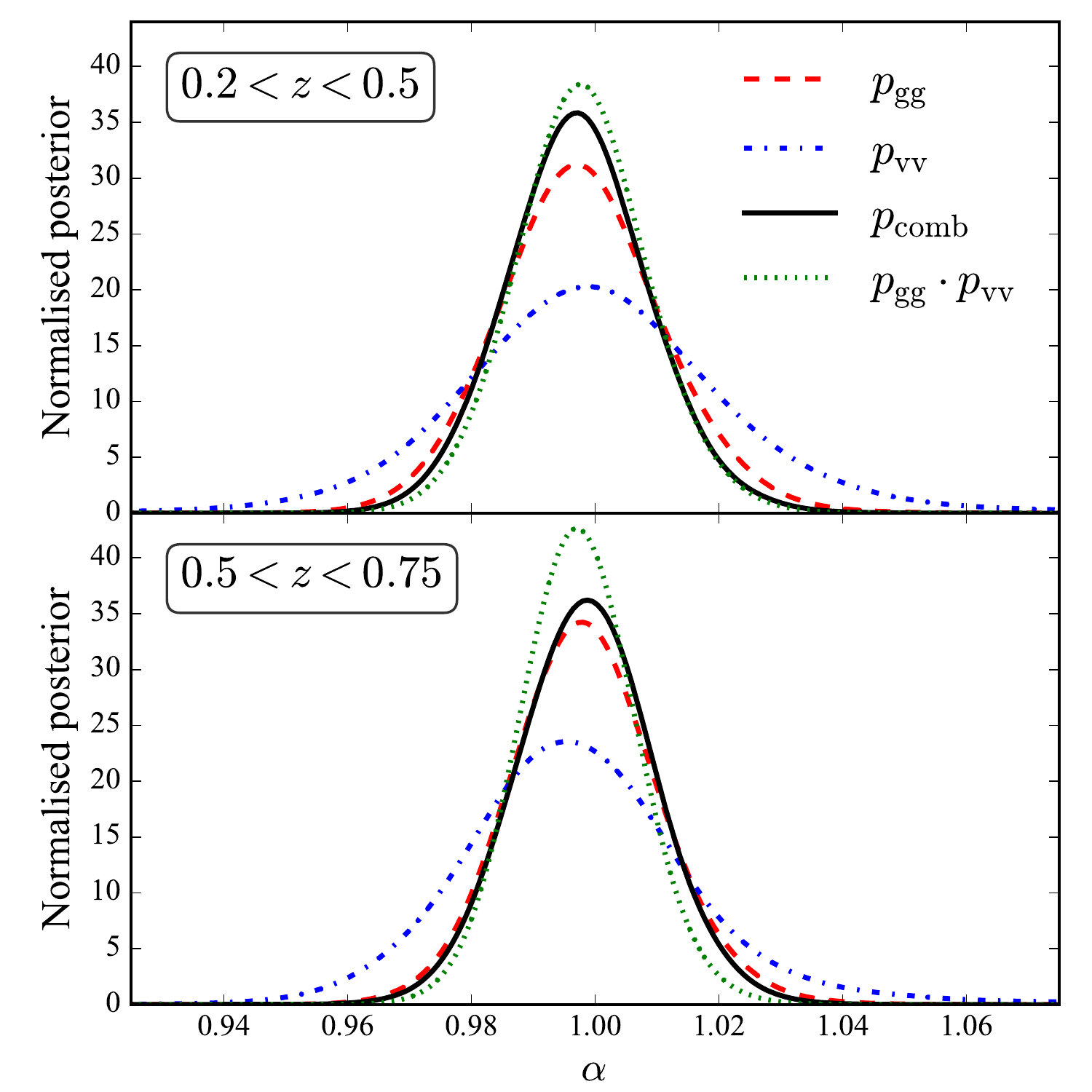}
\caption{The marginal posterior distribution of $\alpha$, from the fits to the mean of 1000 MD-\patchy{} light-cone mocks, of galaxies ($p_{\rm gg}$, red dashed lines), voids ($p_{\rm vv}$, blue dash-dotted lines), and the combined sample with $w=-0.05$ ($p_{\rm comb}$, black solid lines), as well as the joint distribution for galaxies and voids if they were independent ($p_{\rm gg} \cdot p_{\rm vv}$, green dotted lines). The {\it upper} and {\it lower} panels show results from the \lowz{} and \highz{} samples respectively. The distributions are all normalised such that the total area under the curves are 1. The $\sigma_\alpha$ value drawn from the green dotted lines are smaller than the ones for the combined samples listed in Table~\ref{tab:fit_lightcone}. Thus, the 2PCF of galaxies and voids are indeed correlated.}
\label{fig:postterior_patchy}
\end{figure}

To check the correlation between the 2PCF of galaxies and voids, we plot the posterior distribution of $\alpha$, for galaxies, voids, and the combined sample with $w=-0.05$ in Figure~\ref{fig:postterior_patchy}, as well as the one for a joint $\alpha$ constraint if galaxies and voids were completely independent.
However, the $1\,\sigma$ dispersion ($\sigma_\alpha$) of the joint constraint without considering the degeneracies between galaxies and voids are 0.0105 and 0.0094 for the \highz{} and \lowz{} samples respectively.
The values are smaller than those of the combined samples with optimal weights, it indicates that it is crucial to take the correlations into account. This is natural since the cosmic voids considered in this study are constructed from the galaxy distribution.

\subsection{Measurements from BOSS DR12 galaxies and voids}

\begin{figure*}
    \centering
    \includegraphics[width=.9\columnwidth]{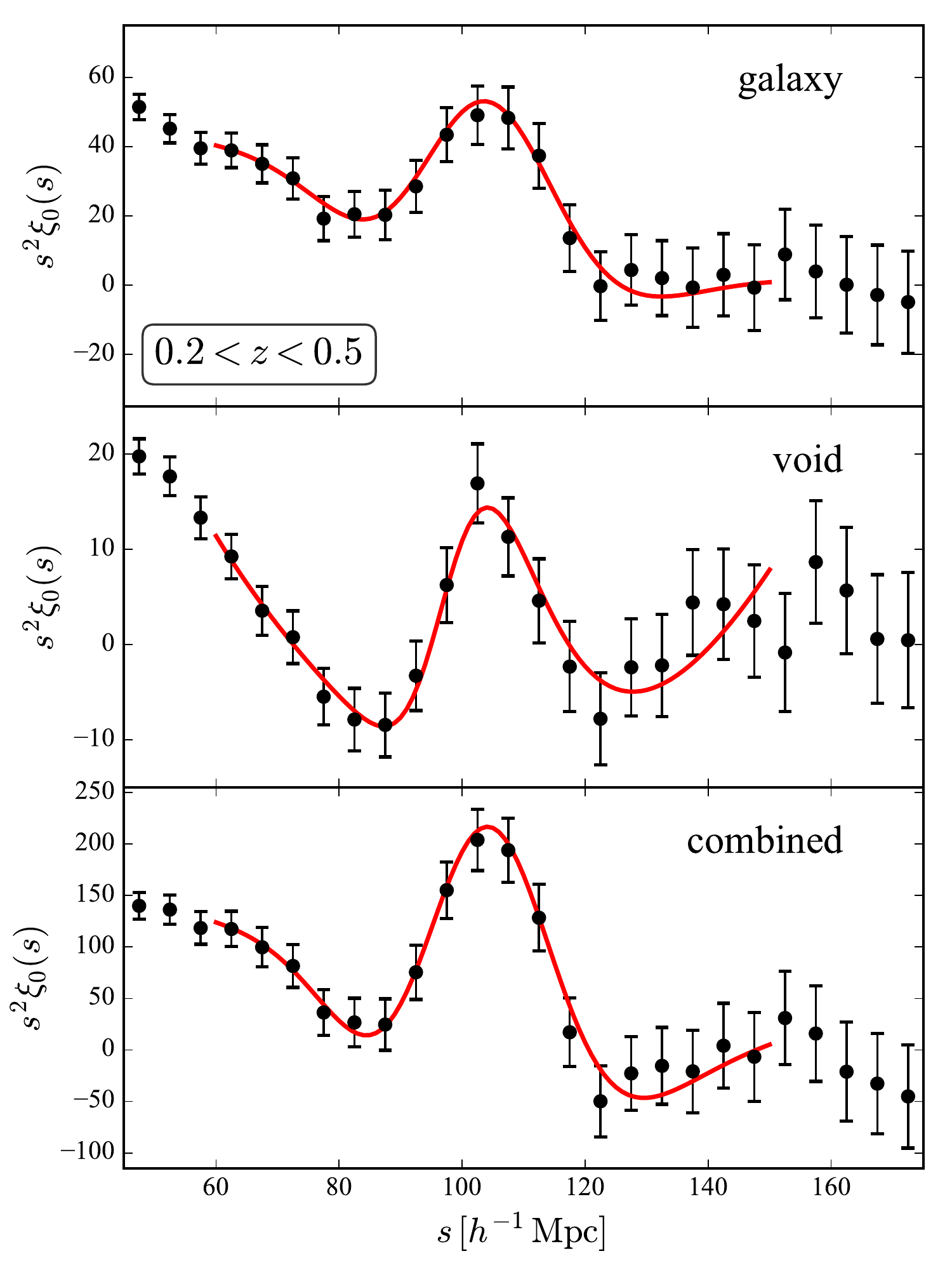}
    \includegraphics[width=.9\columnwidth]{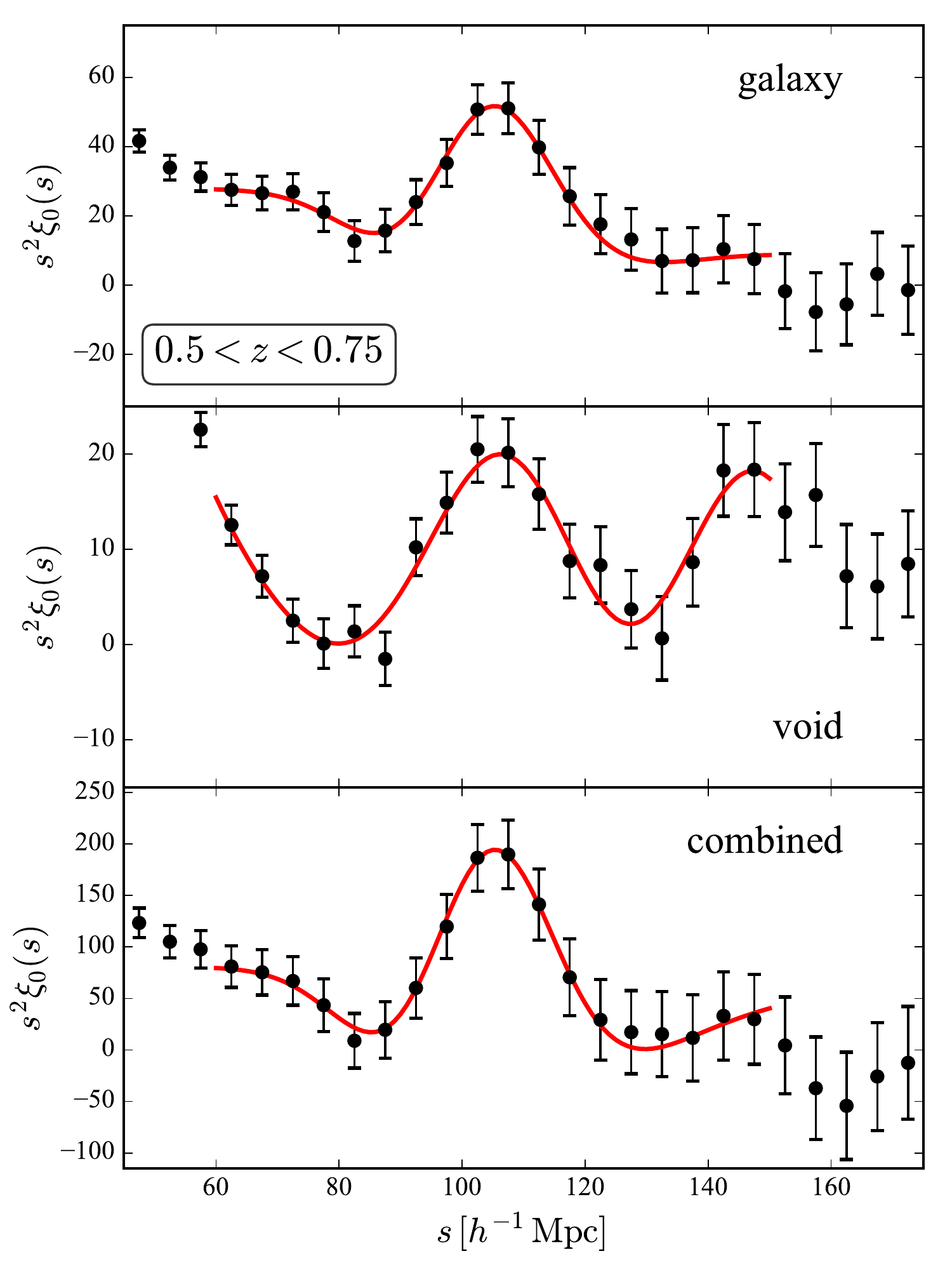}
    \caption{The best-fit (maximum likelihood) curves for the BOSS DR12 data of the galaxy sample, the void sample, and the combined sample with $w=-0.05$, in the redshift bins $0.2 < z < 0.5$ ({\it left}) and $0.5 < z < 0.75$ ({\it right}), respectively.}
    \label{fig:bestfit_data}
\end{figure*}

We now take the optimal void weight obtained from MD-\patchy{} mocks ($w = -0.05$), and apply it to the corresponding BOSS DR12 data. Again we separate the data into \highz{} and \lowz{} bins, and the best-fit (maximum likelihood) curves for the sub-samples are shown in Figure~\ref{fig:bestfit_data}. For all cases the theoretical models are in good agreement with the measured data, indicating robust BAO peak scale measurements.

\begin{figure}
    \centering
    \includegraphics[width=\columnwidth]{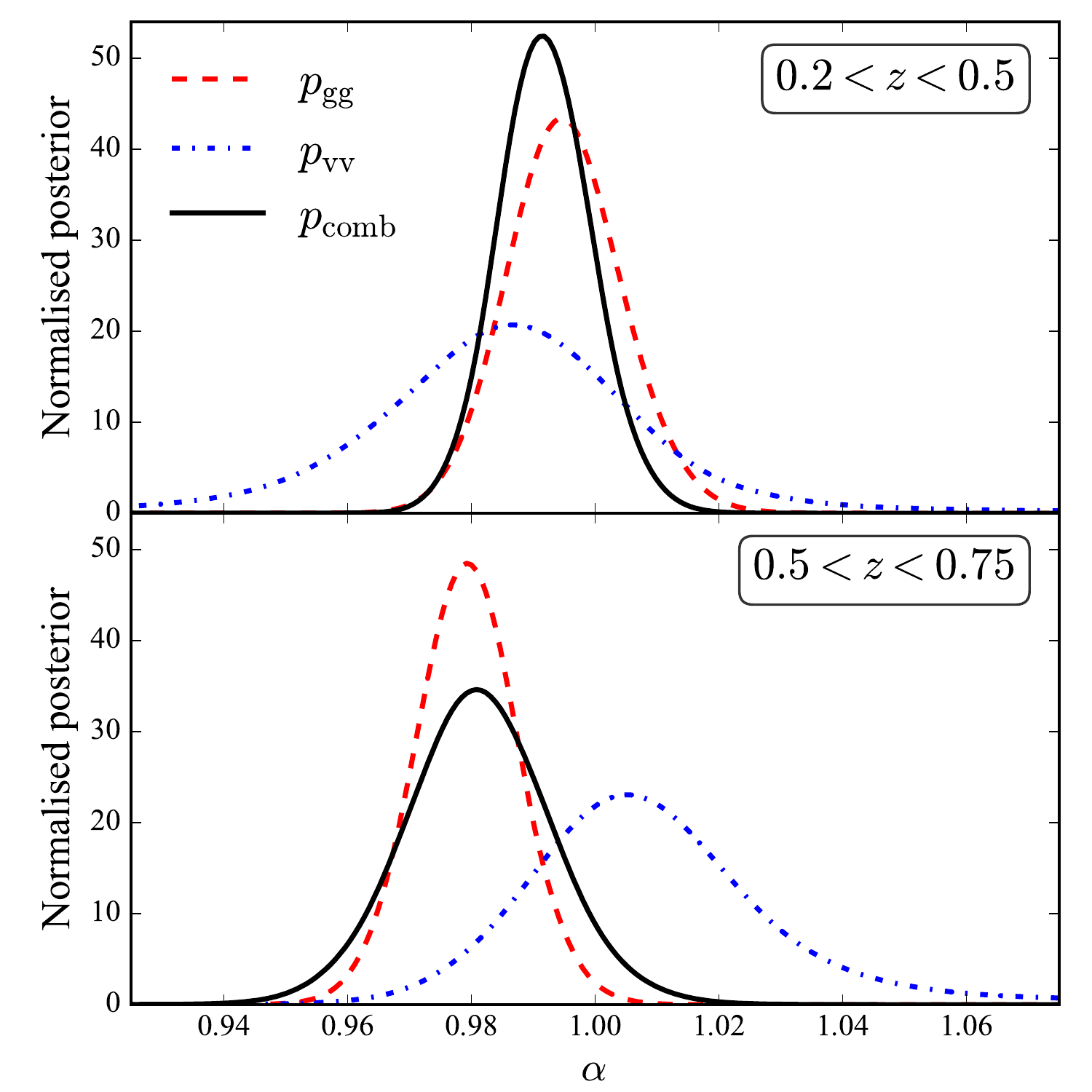}
    \caption{The marginal posterior distribution of $\alpha$, the isotropic BAO peak scale relative to the theoretical prediction from the fiducial cosmology, from fits to the BOSS DR12 data, of galaxies (red dashed lines), voids (blue dash-dotted lines), and the combined sample with $w=-0.05$. The {\it upper} and {\it lower} panels show results from the \lowz{} and \highz{} samples, respectively. And the distributions are normalised such that the total area under the curves are 1.}
    \label{fig:posterior_data}
\end{figure}

\begin{table}
\caption{The fit results for the BOSS DR12 data of the galaxy sample, the void sample, and the combined sample with void weight $w=-0.05$. The degree of freedom for the fit is 11.}
\centering
\begin{tabular}{ccc}
\toprule
 & $0.2 < z < 0.5$ & $0.5 < z < 0.75$ \\
\midrule
& \multicolumn{2}{c}{$\alpha$} \\
galaxy & $0.9951 \pm 0.0092$ & $0.9793 \pm 0.0086$ \\
void & $0.9859 \pm 0.0244$ & $1.0059 \pm 0.0200$ \\
combined & $0.9920 \pm 0.0075$ & $0.9815 \pm 0.0119$ \\
\midrule
& \multicolumn{2}{c}{$\ln{\mathcal{Z}}$} \\
galaxy & $-19.7 \pm 0.17$ & $-13.4 \pm 0.16$ \\
void & $-12.9 \pm 0.08$ & $-16.7 \pm 0.05$ \\
combined & $-22.4 \pm 0.11$ & $-10.7 \pm 0.08$ \\
\midrule
& \multicolumn{2}{c}{$\chi_{\rm min}^2$} \\
galaxy & 18.6 & 5.82 \\
void & 8.15 & 15.5 \\
combined & 23.9 & 3.39 \\
\bottomrule
% & \multicolumn{2}{c}{$\alpha \pm \sigma_\alpha$} \\
% galaxy & $0.9967 \pm 0.0092$ & $0.9801 \pm 0.0094$ \\
% void & $0.9842 \pm 0.0173$ & $0.9979 \pm 0.0315$ \\
% combined & $0.9934 \pm 0.0081$ & $0.9814 \pm 0.0102$ \\
% \midrule
% & \multicolumn{2}{c}{$\chi_{\rm min}^2$} \\
% galaxy & 24.3 & 10.1 \\
% void & 9.21 & 27.8 \\
% combined & 26.8 & 12.0 \\
% \bottomrule
\end{tabular}
\label{tab:fit_data}
\end{table}

The fitted parameters are listed in Table~\ref{tab:fit_data}.
And the $\alpha$ and $\sigma_\alpha$ results for the galaxy samples are again consistent with those in \citet[][]{Vargas2018}.
%The minimum $\chi^2$ for the \lowz{} galaxy sample is however much larger than that of \highz{} galaxies, possibly because we use a different fitting range as the one in \citet[][]{Vargas2018}, and the minimum $\chi^2$ is more sensitive to the fitting range than to either $\alpha$ or $\sigma_\alpha$ (see Appendix~\ref{sec:range}).
Besides, Figure~\ref{fig:posterior_data} shows the posterior distribution of $\alpha$ for the two redshift bins.
For the \lowz{} bin, compared to the galaxy sample, the combined sample shows an 18\,\% improvement on the error of baryon acoustic scale constraint ($( \sigma_{\alpha, {\rm combined}} - \sigma_{\alpha, {\rm galaxy}} ) / \sigma_{\alpha, {\rm galaxy}}$), which is consistent with the expectation value from the results on MD-\patchy{} mocks (13.7\,\%). However, the \highz{} bin does not show improvements with voids. Indeed, the results for the combined sample is even worse.
The next section explains that this result is compatible with statistical fluctuations with present statistics.

\subsection{Statistical fluctuations on the improvement with voids}

\begin{figure*}
    \centering
    \includegraphics[width=\columnwidth]{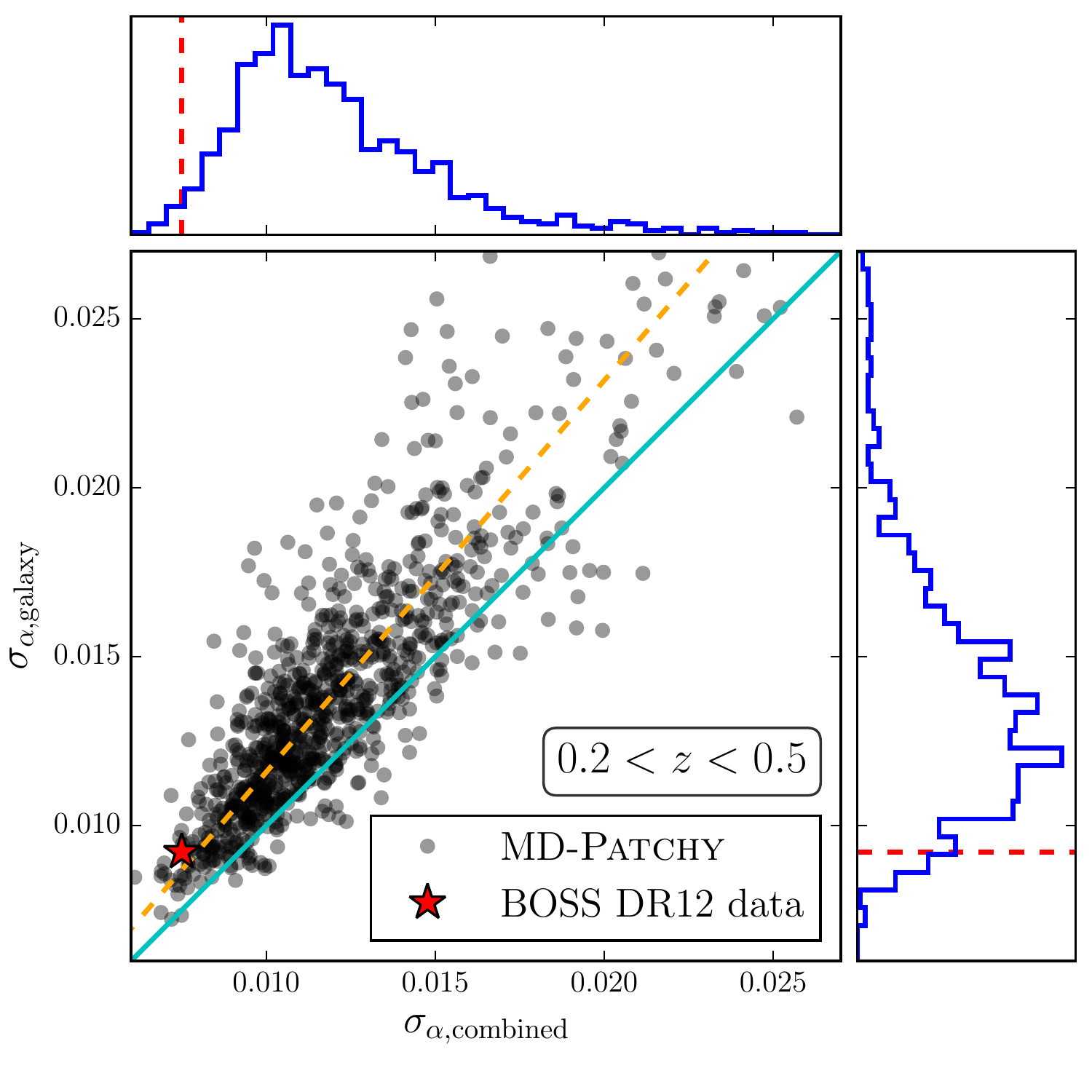}
    \includegraphics[width=\columnwidth]{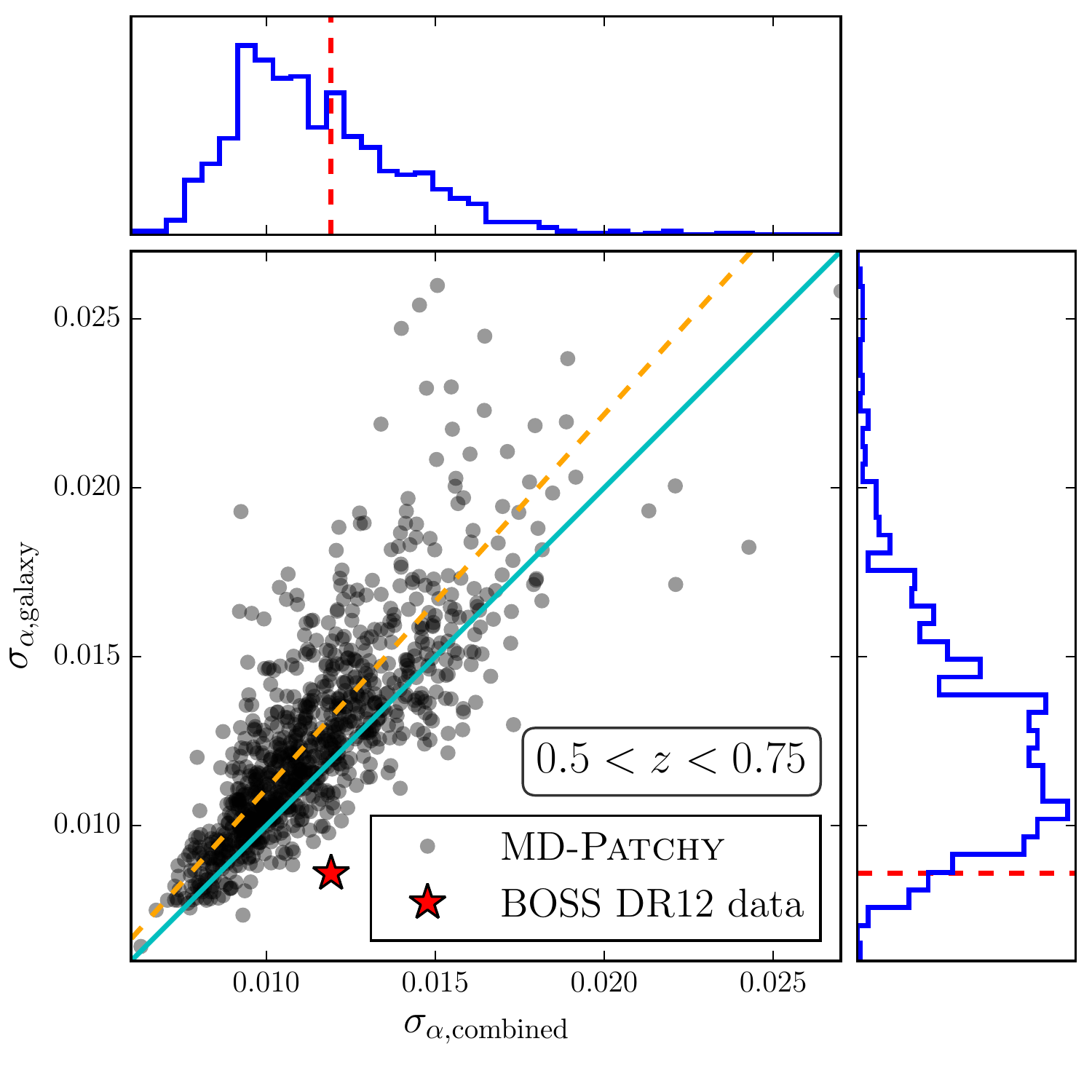}
    \caption{The distribution of the precision of baryon acoustic scale constraint ($\sigma_\alpha$), measured from galaxies and the combined sample with $w=-0.05$ for voids, in the redshift bins $0.2 < z < 0.5$ ({\it left}) and $0.5 < z < 0.75$ ({\it right}), respectively. The dots show results from 1000 individual MD-\patchy{} mocks, and the red stars indicate the measurements from BOSS DR12 data. The cyan solid lines indicate the threshold $\sigma_{\alpha, {\rm combined}} = \sigma_{\alpha, {\rm galaxy}}$, thus realisations with improvements from voids lie above the diagonal lines. And the orange dashed lines show the relative improvement from fits to the mean of the mocks, i.e. $\sigma_{\alpha, {\rm combined}} = 0.863 \, \sigma_{\alpha, {\rm galaxy}}$ (13.7\,\% improvement) for the \lowz{} sample and $\sigma_{\alpha, {\rm combined}} = 0.902 \, \sigma_{\alpha, {\rm galaxy}}$ (9.8\,\% improvement) for the \highz{} case. Furthermore, the histograms show the distribution of $\sigma_\alpha$ obtained from individual mocks, and red dashed lines indicate the fitted values for the BOSS data.}
    \label{fig:sigma_single}
\end{figure*}

In this section, we measure the BAO scale constraints from the 1000 individual mocks to study the distribution of the improvement on baryon acoustic scale constraint.
For each mock realisation, the fitted $\sigma_\alpha$ for galaxies ($\sigma_{\alpha, {\rm galaxy}}$) and the combined sample with $w=-0.05$ ($\sigma_{\alpha, {\rm combined}}$) are shown in Figure~\ref{fig:sigma_single}.
Indeed for the majority of the mocks, $\sigma_\alpha$ is reduced by including voids. And the distributions are consistent with the results from fits to the mean 2PCF of the mocks (blue dashed lines in Figure~\ref{fig:sigma_single}).
In particular, the number of mocks with $\sigma_{\alpha, {\rm combined}} < \sigma_{\alpha, {\rm galaxy}}$ are 868 and 742 for the \lowz{} and \highz{} samples respectively, indicating that the chance of having no contribution from voids is $\sim 25\,\%$.

%Denoting $\sigma_{\alpha, {\rm galaxy}}$ as the fitted error on $\alpha$ for the galaxy sample, and $\sigma_{\alpha, {\rm combined}}$ as the result from the combined sample with the void weight of $w=-0.05$, we define the relative improvement on the constraint as $( \sigma_{\alpha, {\rm combined}} - \sigma_{\alpha, {\rm galaxy}} ) / \sigma_{\alpha, {\rm galaxy}}$. In this case, a negative value indicates that the BAO scale determination is better.
%We plot the distribution of the relative improvement on the BAO scale constraint from the 1000 MD-\patchy{} mocks in Figure~\ref{fig:diff_sigma_highz} and Figure~\ref{fig:diff_sigma_lowz} for the \highz{} and \lowz{} redshift bins respectively.
%There are single peaks for both distributions, that are located in negative regions, indicating that on average there are improvements with voids. However, a considerable fraction of the mocks does not show better BAO scale constraint, with some realisations showing even much larger $\sigma_{\alpha, {\rm combined}}$. Therefore, for a single realisation, there are chances that voids have no contribution to the BAO measurement.

The fitted $\sigma_{\alpha, {\rm galaxy}}$ and $\sigma_{\alpha, {\rm combined}}$ for the BOSS DR12 data are not far away from the most probable values indicated by the distribution of results from individual mocks.
Thus, the failure of improvement with voids for the \highz{} bin of BOSS DR12 data is consistent with statistical fluctuations (cosmic variance).

\begin{figure}
    \centering
    \includegraphics[width=\columnwidth]{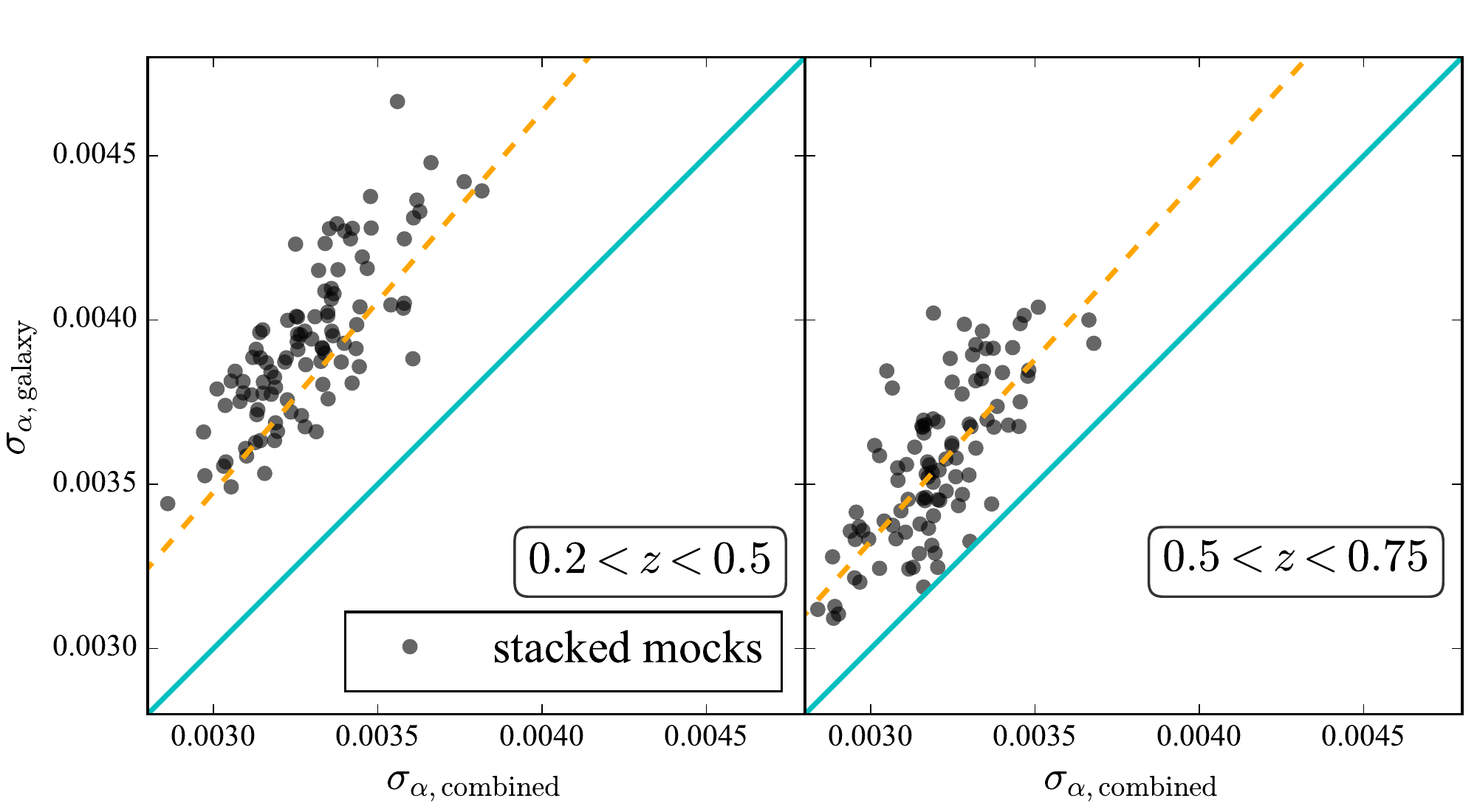}
    \caption{The distribution of $\sigma_\alpha$ measured from galaxies and and combined sample with $w=-0.05$ for voids, obtained from fits to the mean of every 10 MD-\patchy{} mocks, in the redshift bins $0.2 < z < 0.5$ ({\it upper}) and $0.5 < z < 0.75$ ({\it lower}), respectively. The cyan solid lines indicate the threshold that there is no improvement by including voids, while orange dashed lines show the relative improvement from fits to the mean of all mocks (13.7\,\% for \lowz{} and 9.8\,\% for \highz{}). Thus, for all the 100 combined realisations in both redshift bins, there are better BAO scale constraints with voids.}
    \label{fig:sigma_stack}
\end{figure}

We further explore the performance of our method for future surveys with larger volumes (smaller statistical fluctuations) by taking the mean 2PCF of every 10 mocks, to have a larger effective volume for each 2PCF.
We dub this set of mocks `stacked mocks'. With the 100 realisations of 2PCFs, we then evaluate the covariance matrix, and apply the modified de-wiggled model to obtain the BAO scale constraints for galaxies and the combined sample with $w=-0.05$.
The results are shown in Figure~\ref{fig:sigma_stack}.
In this case for all of the realisations, we have always a better constraint on $\alpha$ by including voids.
This indicates that our method is more promising for a larger sample size. 

%%%%%%%%%%%%%%%%%%%%%%

\section{Conclusions}
\label{sec:con}
%In previous works, we presented the 2-point clustering studies on a novel cosmic void definition through the Delaunay Triangulation technique, with a cosmological independent and efficient void finding code \citep[\textsc{dive},][]{Zhao2016}. The voids defined in this way are circumspheres of Delaunay cells constructed using haloes or galaxies. We perform a radius selection and keep only large voids with radii above a threshold, to remove the contamination on the BAO signal of voids-in-clouds. This radius threshold, the only parameter for our void definition, is chosen to maximise the signal-to-noise ratio of the BAO peak from mocks that are used to estimate the covariance matrices of a given set of observational data \citep[][]{Liang2016}. Then, we apply this radius threshold ($16\,\hmpc$) to voids resolved from the given set of data, and have evidenced a 3.2-$\sigma$ BAO peak from BOSS DR11 data \citep[][]{Kitaura2016}.

In this paper, we focus on the BAO peak position constraints from void clustering statistics, and the consequences when combined with halo/galaxy data on the determination of cosmological parameters. Our voids are defined as the circumspheres of Delaunay tetrahedra constructed using haloes or galaxies. They have high overlapping fraction, and should not be considered as individual spherical under-dense regions, but tracers of the underlying cosmic web structures. Furthermore, a radius selection of $R_{\rm V} \ge 16\,\hmpc$ is applied to identify voids only tracing under-densities.

We first perform a theoretical and ideal study on void BAO using 100 realisations of cubic \patchy{} mocks. To reduce the statistical fluctuations induced by a single realisation, we apply the analyses to the mean of 2PCFs from the 100 mocks, with the error being at one realisation level.
From this test, we find that the typical de-wiggled BAO model does not fit the 2PCF of voids well. To circumvent this problem, we have modified the de-wiggled model by including an extra parameter to describe the non-wiggle power spectrum of voids. This new model still fits worse for voids than for haloes, but the goodness of fits are comparable. And it is much better than the typical de-wiggled model for voids.

With the modified BAO model, we find the BAO peak position constraint from voids weaker, but close to that from haloes without performing BAO reconstruction. This can be explained by the goodness of fit, as well as the fact that the signal-to-noise ratio of void 2PCF is lower than that of the reference haloes/galaxies \citep[][]{Kitaura2016}. Furthermore, we have evidenced an outward BAO position shift for voids, this is consistent with the fact that over-dense regions expand, and yield outward BAO peak motions.

We then apply the BAO reconstruction method \citep[][]{Eisenstein2007, Padmanabhan2012} to the haloes from \patchy{} cubic mocks, with a smoothing length of $5\,\hmpc$. In this way we find the BAO peak position constraint from haloes improved by $\sim 25\,\%$. Besides, the BAO shift of haloes is corrected. Later on, we apply our void finder to the reconstructed haloes, and construct in this way a post-reconstruction void sample.
Even though the BAO shift of voids is not corrected, there is a $\sim 20\,\%$ reduction of the error on the BAO scale constraint.
The BAO position of haloes and voids are more consistent than in the pre-reconstruction case.

This consistency allows us to perform a joint constraint using both haloes and voids, since we expect zero BAO shifts for both haloes and voids for an ideal reconstruction.
To this end, we use 1000 realisations of post-reconstruction MultiDark \patchy{} DR12 mocks, which are calibrated to mimic the BOSS DR12 observational data.
Since the BAO peak position of haloes in the cubic mocks is more robust (the measured $\alpha - 1$ in Table~\ref{tab:fit_new_model} is closer to 0 with smaller uncertainty for haloes, given the input cosmology for generating the mocks as the fiducial cosmology for the fits), we treat galaxies as our basis, and study the contribution of voids.
Thus, we apply a single weight $w$ to the whole void sample, and vary it from $-1$ to $1$ to combine galaxies with voids, and study the BAO position constraint of the joint sample.
In this way, we find an optimal void weight of $w=-0.05$ for both \lowz{} and \highz{} redshift bins. With the weight near this value, we minimize the fitted error of BAO peak position, without biasing it.

The sign of the $w$ is consistent with the fact that the bias of haloes and voids have opposite signs, and the negative weight promotes the contribution of the negative cross correlations. Besides, the small value of weight implies that the net additional information from voids is small (but not 0). The non-zero contribution of voids can be explained by the information contained in under-dense regions, while it is missing in the 2PCF of a halo/galaxy sample.
Since BAO reconstruction transfer information encoded in high order statistics of the density field to the 2PCF, we expect the improvement from voids to be reduced in the future, with better reconstruction methods.
However, due to the lack of information from under-densities, there should always be some cosmological gain from voids even with an ideally reconstructed galaxy sample.
We will check if improved reconstruction algorithms \citep[e.g.][and Kitaura et al., in preparation]{Hada2018, Sarpa2019} would reduce the contribution from voids in a future work.

Moreover, after applying this optimal weight to the MultiDark \patchy{} DR12 mocks, we find $13.7\,\%$ and $9.8\,\%$ improvement for the error on the BAO peak position over that from the galaxy sample alone, for the \lowz{} and \highz{} samples respectively.
We further divide the 1000 mocks to 2 independent sets of 500 realisations, and observe the same level of improvements. This ensures that the improvements are indeed from our void sample, rather than statistical fluctuations.
These levels of improvements are equivalent to enlarging over $20\,\%$ of the sample size, while we do not require any extra inputs other than those for galaxy BAO analyses.
In addition, the cosmological gain is actually smaller than the extreme case assuming no correlations between galaxies and voids, which indicates that the degeneracy has to be taken into account.
In fact, we have included the degeneracy between galaxies and voids, and especially for overlapping voids, by using the full covariance from the mocks coherently.

We finally apply the same void weight scheme to the BOSS DR12 data, and find an $18\,\%$ improvement on the BAO position constraint for the \lowz{} data. However, we do not observe any improvement for the \highz{} data, and the constraint from the joint sample is even worse. By studying the performance of our method for the 1000 individual mocks, we find the influence of cosmic variance not negligible. And the failure of improvement for the \highz{} sample is consistent with statistical fluctuations for a single realisation.
Indeed, for the \highz{} sample, the probability of having no improvement is as large as $\sim 25\,\%$, as only $\sim 75\,\%$ of the 1000 mocks show better BAO scale constraint with voids.
However, the improvement is more robust for a larger sample size.
%In the ideal case (i.e., the sample size is sufficient large), the significance of a better constraint using voids can be $> 5\,\sigma$.

It is worth noting that there are only 2 free parameters in our method, the optimal radius threshold for selecting voids, and the weight applied to the full void population. For changes to the galaxy sample or fiducial cosmology, these 2 parameters should be re-calibrated using a large set of realistic mocks.

We leave the application to cosmological parameter constraints using our method to a future paper. Besides, a better radius selection criteria for post-reconstruction voids as well as more detailed weighting schemes for combining galaxies and voids will be investigated in forthcoming works.
Finally, we wish to underline that the analysis in this work is done within the framework of $\Lambda$CDM cosmology, and quantitative results could be different for different Dark Energy or modified gravity models.

%%%%%%%%%%%%%%%%%%%%%%%%%%%%

\section*{acknowledgments}
We thank Prof. Houjun Mo for useful discussions. CZ, YL, and CT are supported by Tsinghua University with a 985 grant, 973 programme 2013CB834906, NSFC grant No. 11033003 and 11173017, and sino french CNRS-CAS international laboratories LIA Origins and FCPPL. CZ also acknowledges supports from NSFC grant No. 11673025 and a Royal Society Newton Advanced Fellowship.
FSK thanks support from the grants RYC2015--18693 and AYA2017-89891-P.
GY acknowledges financial  support  from  the {\it Ministerio de Econom\'ia y Competitividad} and the {\it Fondo Europeo de Desarrollo Regional} (MINECO/FEDER, UE) in Spain through grant AYA2015-63810-P.
MVM is partially supported by Programa de Apoyo a Proyectos de Investigaci\'on e Innovaci\'on Tecnol\'ogica (PAPITT) No. IA102516, Proyecto Conacyt Fronteras No. 281 and from Proyecto LANCAD-UNAM-DGTIC-319.

The computations have been performed on the \textsc{MareNostrum} supercomputer at the Barcelona Supercomputing Centre, thanks to the computing time awarded by Red Espa\~nola de Supercomputaci\'on.
%granted by the project \textit{The Marenostrum Numerical Cosmology Project: Grand Challenge simulations of structure formation in the Universe} leaded by Gustavo Yepes Alonso,
We also acknowledge the HPC facilities \textsc{BeiLuo} at Tsinghua University and \textsc{Erebos}/\textsc{Theia}/\textsc{Geras} at Leibniz-Institut f\"{u}r Astrophysik Potsdam for some of the computations.

Funding for SDSS-III has been provided by the Alfred P. Sloan Foundation, the Participating Institutions, the National Science Foundation, and the U.S. Department of Energy Office of Science. The SDSS-III web site is \url{http://www.sdss3.org/}.

SDSS-III is managed by the Astrophysical Research Consortium for the Participating Institutions of the SDSS-III Collaboration including the University of Arizona, the Brazilian Participation Group, Brookhaven National Laboratory, Carnegie Mellon University, University of Florida, the French Participation Group, the German Participation Group, Harvard University, the Instituto de Astrofisica de Canarias, the Michigan State/Notre Dame/JINA Participation Group, Johns Hopkins University, Lawrence Berkeley National Laboratory, Max Planck Institute for Astrophysics, Max Planck Institute for Extraterrestrial Physics, New Mexico State University, New York University, Ohio State University, Pennsylvania State University, University of Portsmouth, Princeton University, the Spanish Participation Group, University of Tokyo, University of Utah, Vanderbilt University, University of Virginia, University of Washington, and Yale University.

%%%%%%%%%%%%%%%%%%%%%%%%%%%%%%%%%%%%%%%%%%%%%%%%%%

%%%%%%%%%%%%%%%%%%%% REFERENCES %%%%%%%%%%%%%%%%%%

% The best way to enter references is to use BibTeX:

\bibliographystyle{mnras}
\bibliography{VoidBAO} % if your bibtex file is called example.bib

% Alternatively you could enter them by hand, like this:
% This method is tedious and prone to error if you have lots of references
%\begin{thebibliography}{99}
%\bibitem[\protect\citeauthoryear{Author}{2012}]{Author2012}
%\bibitem[\protect\citeauthoryear{Others}{2013}]{Others2013}
%Others S., 2012, Journal of Interesting Stuff, 17, 198
%\end{thebibliography}

%%%%%%%%%%%%%%%%%%%%%%%%%%%%%%%%%%%%%%%%%%%%%%%%%%

%%%%%%%%%%%%%%%%% APPENDICES %%%%%%%%%%%%%%%%%%%%%

\appendix

\section{Fitting range}
\label{sec:range}
\begin{figure*}
    \centering
    \includegraphics[width=0.7\textwidth]{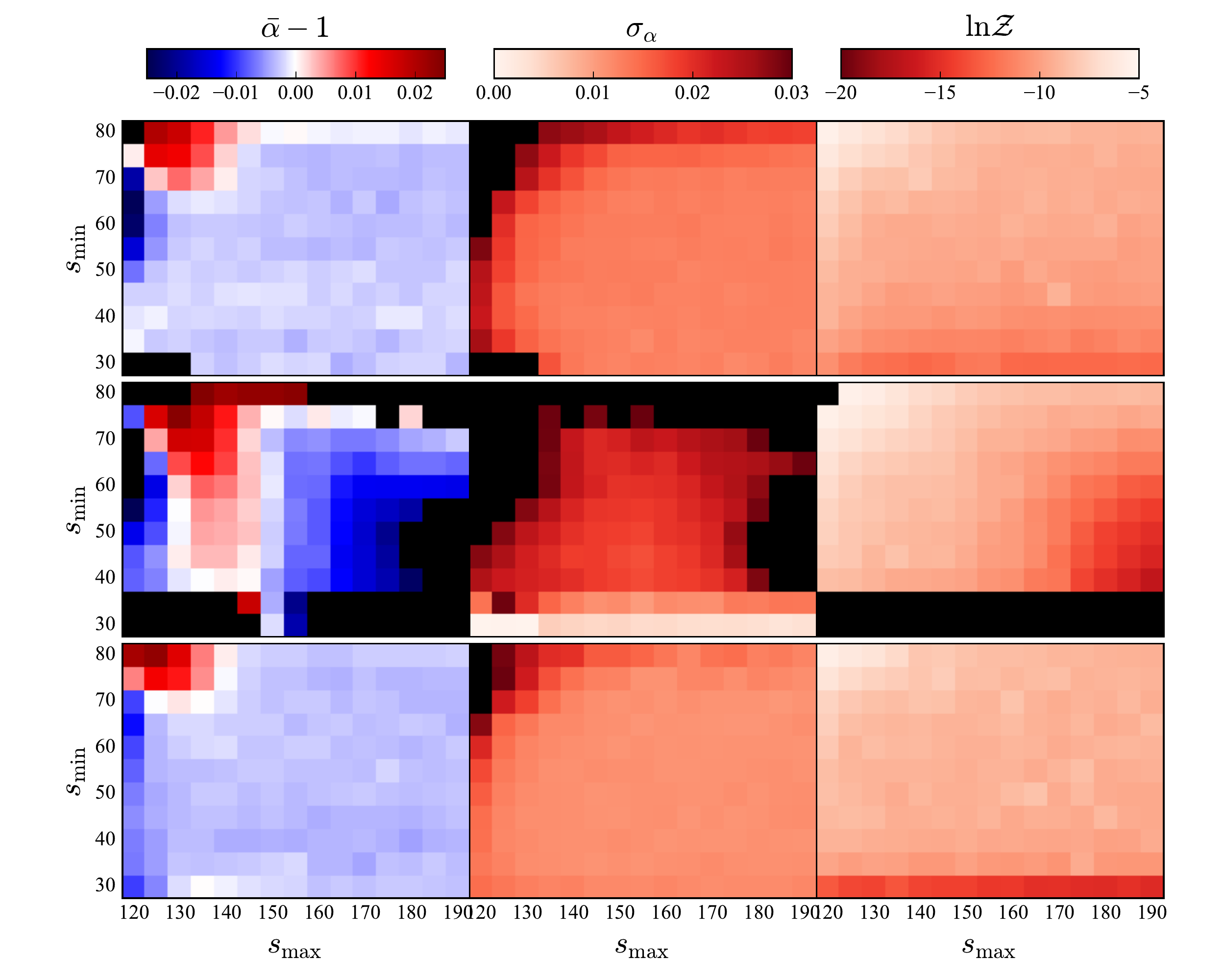}
    \caption{The BAO fit results with different fitting ranges for the mean of 1000 MD-\patchy{} light-cone mocks in redshift bin $0.2 < z < 0.5$.}
    \label{fig:range_lowz}
\end{figure*}

\begin{figure*}
    \centering
    \includegraphics[width=0.7\textwidth]{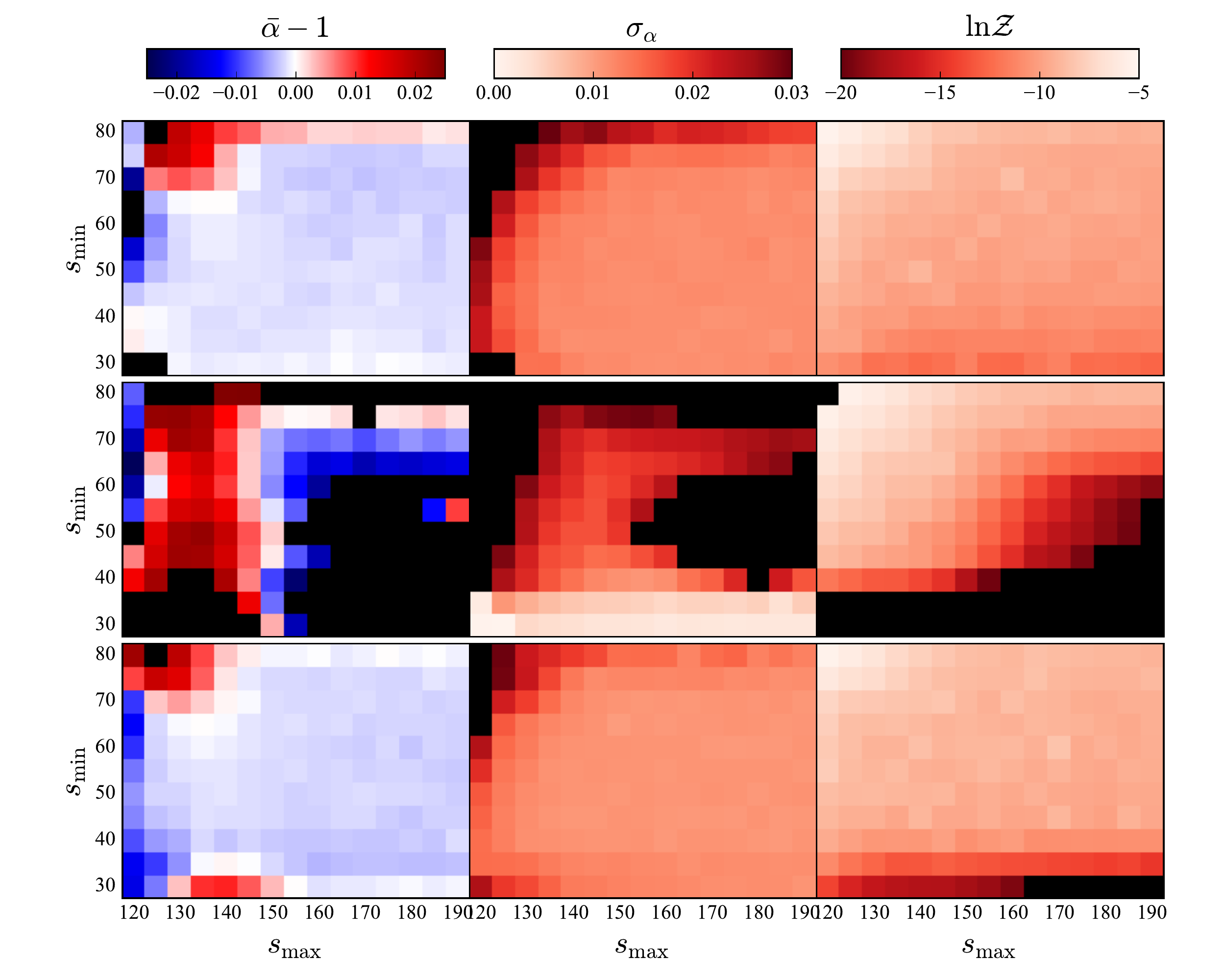}
    \caption{The BAO fit results with different fitting ranges for the mean of 1000 MD-\patchy{} light-cone mocks in the redshift bin $0.5 < z < 0.75$.}
    \label{fig:range_highz}
\end{figure*}

To explore the impact of the fitting range on the baryon acoustic scale constraints, we vary the lower bound in the range $s_{\rm min} \in [30, 80]\,\hmpc$ and the upper bound in range $s_{\rm max} \in [120, 190]\,\hmpc$ with the bin size of $5\,\hmpc$ for both cases. We then apply our BAO fitting procedure to the mean of 1000 MD-\patchy{} light-cone mocks. The fit results, including $\alpha$, $\sigma_\alpha$, and the Bayesian evidence, of the galaxy sample, the void sample, and the combined sample with the void weight $w=-0.05$ are shown in Figure~\ref{fig:range_highz} and Figure~\ref{fig:range_lowz} for the \highz{} and \lowz{} redshift bins respectively.

%It can be seen that the fitting results for voids are very sensitive to the fitting range. However, the results for the galaxy sample and the combined sample are both robust for $s_{\rm min} \in [50, 65]\,\hmpc$ and $s_{\rm max} \in [160, 190]\,\hmpc$.
The fit results for the galaxy sample and the combined sample are both robust for $s_{\rm min} \in [40, 65]\,\hmpc$ and $s_{\rm max} \in [140, 190]\,\hmpc$. However, the results for voids are very sensitive to the fitting range. Moreover, it is not possible to find the optimal fitting range for all the 3 cases. As the main goal of this paper is to investigate whether the combined sample yields better BAO scale constraint than the galaxy sample alone, we finally choose the fitting range $s \in [60, 150]\,\hmpc$ in this work.
In this case, the fit results for galaxies and the combined sample are both insensitive to the range, and there is a sufficient number of data points in this range.
The results for voids are however not optimal for this choice, but they are not strongly affected, given the fact that the fitted alpha is only biased by $< 2\,\%$, comparable to the constraining power.
Indeed, we have also checked the optimal weight for several different fitting ranges near the one we choose, and find it not varying. This further confirms that our results are robust with respect to the fitting range.

%%%%%%%%%%%%%%%%%%%%%%%%%%%%%%%%%%%%%%%%%%%%%%%%%%

% Don't change these lines
\bsp	% typesetting comment
\label{lastpage}
\end{document}